\documentclass[aps,pra,twocolumn,floats,amsmath,amssymb,superscriptaddress]{revtex4-1}
\usepackage[utf8]{inputenc}
\usepackage{graphicx}
\usepackage{epsfig}
\usepackage{amsfonts}
\usepackage{amsmath}
\usepackage{natbib}
\usepackage{multirow}
\usepackage{bm}
\usepackage{epstopdf}
\usepackage{ulem}
\usepackage{color}
\usepackage{hyperref}

\definecolor{green}{rgb}{0, 0.75, 0}

\newcommand{\exciting}{{\usefont{T1}{lmtt}{b}{n}exciting}}
\begin{document}
\title{
A consistent picture of excitations in cubic BaSnO$_{3}$ revealed by combining theory and experiment}
\author{Wahib Aggoune}
\email{aggoune@physik.hu-berlin.de}
\affiliation{Institut f\"{u}r Physik and IRIS Adlershof, Humboldt-Universit\"{a}t zu Berlin, 12489 Berlin, Germany}
\affiliation{Fritz-Haber-Institut der Max-Planck-Gesellschaft, 14195 Berlin, Germany}
\author{Alberto Eljarrat}
\email{alberto.eljarrat@physik.hu-berlin.de}
\affiliation{Institut f\"{u}r Physik and IRIS Adlershof, Humboldt-Universit\"{a}t zu Berlin, 12489 Berlin, Germany}
\author{Dmitrii Nabok}
\affiliation{Institut f\"{u}r Physik and IRIS Adlershof, Humboldt-Universit\"{a}t zu Berlin, 12489 Berlin, Germany}
\affiliation{European Theoretical Spectroscopic Facility (ETSF)}
\author{Klaus Irmscher}
\affiliation{Leibniz-Institut f\"{u}r Kristallz\"{u}chtung, 12489 Berlin, Germany}
\author{Martina Zupancic}
\affiliation{Leibniz-Institut f\"{u}r Kristallz\"{u}chtung, 12489 Berlin, Germany}
\author{Zbigniew Galazka}
\affiliation{Leibniz-Institut f\"{u}r Kristallz\"{u}chtung, 12489 Berlin, Germany}
\author{Martin Albrecht}
\affiliation{Leibniz-Institut f\"{u}r Kristallz\"{u}chtung, 12489 Berlin, Germany}
\author{Christoph Koch}
\affiliation{Institut f\"{u}r Physik and IRIS Adlershof, Humboldt-Universit\"{a}t zu Berlin, 12489 Berlin, Germany}\author{Claudia Draxl}
\affiliation{Institut f\"{u}r Physik and IRIS Adlershof, Humboldt-Universit\"{a}t zu Berlin, 12489 Berlin, Germany}
\affiliation{Fritz-Haber-Institut der Max-Planck-Gesellschaft, 14195 Berlin, Germany}
\affiliation{European Theoretical Spectroscopic Facility (ETSF)}

\date{\today }
\begin{abstract}
\section*{Abstract}

Among the transparent conducting oxides, the perovskite barium stannate is most promising for various electronic applications due to its outstanding carrier mobility achieved at room temperature. However, most of its important characteristics, such as band gaps, effective masses, and absorption edge, remain controversial. Here, we provide a fully consistent picture by combining state-of-the-art {\it ab initio} methodology with forefront electron energy-loss spectroscopy and optical absorption measurements. Valence electron energy-loss spectra, featuring signals originating from band gap transitions, are acquired on defect-free sample regions of a BaSnO$_{3}$ single crystal. These high-energy-resolution measurements are able to capture also very weak excitations below the optical gap, attributed to indirect transitions. By temperature-dependent optical absorption measurements, we assess band-gap renormalization effects induced by electron-phonon coupling. Overall, we find for the effective electronic mass, the direct and the indirect gap, the optical gap, as well as the absorption onsets and spectra, excellent agreement between both experimental techniques and the theoretical many-body results, supporting also the picture of a phonon-mediated mechanism where indirect transitions are activated by phonon-induced symmetry lowering. This work demonstrates a fruitful connection between different high-level theoretical and experimental methods for exploring the characteristics of advanced materials.

\end{abstract}
\maketitle
\section*{Introduction}
Transparent conducting oxides (TCO) have attracted attention of researchers and engineers due to their wide range of potential applications. Among them, the perovskite BaSnO$_{3}$, when doped with lanthanum, has turned out as a most promising candidate for the next generation of electronic devices. This is mainly due to the combination of an extraordinary room-temperature mobility, reaching 320 cm$^{2}$V$^{-1}$s$^{-1}$ -- which is the highest ever measured in TCOs -- and a significant degree of transparency in the visible range~\cite{Hkim+12ape}. Additionally, BaSnO$_{3}$ exhibits excellent thermal stability. As such, it is already explored as a channel material in field-effect transistors and as an electron-transporting layer for a highly efficient dye-sensitized solar cells~\cite{shin+13acsn,Hkim+12ape,Hkim+12prb,Useong+15apl,shin+16apl,zhu+17jmca}. The high mobility in this material is attributed to the low effective electron mass of the conduction band, composed of Sn-\textit{s} states, as well as to the reduced carrier scattering due the large dielectric constant~\cite{Hkim+12prb,Niedermeier+17aps}. However, at higher doping densities, it was shown that ionized-impurity scattering may limit the mobility \cite{krish+17prb}. As an alternative to La-doping, an exciting perspective has been opened by polar-discontinuity doping, {\it i.e.}, by forming interfaces or heterostructures of the nonpolar BaSnO$_{3}$ with a polar oxide perovskite. In this case, a high-mobility two-dimensional electron gas can be formed in BaSnO$_{3}$ without doping~\cite{lee+17armr,krish+17prb,Paudel+17prb,Krish+16apl,Useong+15apl,kim+19sp,Martina+20prm,agg+21npj}. 

Detailed knowledge of the pristine material is prerequisite for the fundamental understanding of its doped or nanostructured counterparts and, consequently, for developing such promising applications. In the last years, the electronic and optical properties of BaSnO$_{3}$ have been intensively investigated from both theory and experiment~\cite{mizoguchi+04jacs,hadjarab+07apdap,Hkim+12prb,luo+12apl,
Stanislavchuk+12jap,Kang+18apl,dongmin+14apl,chambers+16apl,galazka+16jpcm,yan+18jvsta,rylan+20jcg,Krish+16apl,Dabaghmanesh+13jpcm,Scanlon+13prb,kim+13jssc,krish+17prb,monserrat+18prb}. However, effective masses, band gaps, and absorption edge are still controversial. While, the value of the fundamental gap is under debate, there is evidence of its indirect character, obtained by angle-resolved photoemission spectroscopy~\cite{beom+17cap} and absorption measurements~\cite{galazka+16jpcm}. Measured values ~\cite{mizoguchi+04jacs,hadjarab+07apdap,Hkim+12prb,luo+12apl,Stanislavchuk+12jap,dongmin+14apl,chambers+16apl,galazka+16jpcm,yan+18jvsta,rylan+20jcg} for the indirect fundamental gap vary between 2.85 and 3.3 eV, while for the direct counterpart between 3.1 and 4 eV, with a majority of the latter being around 3.5 eV. The corresponding estimates of the difference between the two range between 0.15 eV and 0.5 eV. These diverse findings were assigned to the presence of uncontrolled defects as well as to differences in measurement techniques. Also the experimental estimates of the effective electron mass are scattered, ranging between 0.19 and 0.65 m$_{\mathrm{0}}$~\cite{Hkim+12prb,dongmin+14apl,james+16apl,Niedermeier+17aps,rylan+20jcg}. 

From the theoretical side, density-functional-theory (DFT) calculations based on semi-local exchange-correlation (xc) functionals~\cite{moreira+12jap} find an indirect fundamental gap, that is underestimated by about 75\%. Incorporating 25\% of Hartree-Fock exchange, the hybrid functional HSE06~\cite{HSE+06jcp} naturally improves the results. The still too small gap (by about 15\%) was explained by the significant dielectric screening included in the functional~\cite{Krish+16apl,Dabaghmanesh+13jpcm,Scanlon+13prb,kim+13jssc,krish+17prb}. The calculated electron masses also widely scatter between 0.06 m$_{\mathrm{0}}$ and 0.9 m$_{\mathrm{0}}$, depending on the xc-treatment~\cite{dongmin+14apl,kim+13jssc}.

Absorption measurements show a weak signal at the onset \cite{Hkim+12prb}, suggesting it to be related to the indirect gap. This was later confirmed \cite{galazka+16jpcm}, and a temperature-induced shift about -0.18 eV was reported, going from 5 to 297 K. In addition, a recent theoretical study~\cite{monserrat+18prb} reveals phonon-assisted absorption across the indirect gap, also leading to the temperature dependence of the spectrum. 

For an in-depth understanding of this puzzle, several parameters need to be disentangled: These are the values of the fundamental direct and indirect gaps, the impact of e-ph coupling in terms of zero-point vibrations (ZPV) and temperature, as well as electron-hole (e-h) interaction (exciton binding energies) for direct as well as indirect transitions. 
To this end, we present a combined theoretical and experimental study of the electronic structure and the optical excitations of cubic BaSnO$_{3}$.

Our theoretical analysis is based on many-body perturbation theory (MBPT) in the framework of the $G_0W_0$ approximation~\cite{hedi65pr,hybe-loui85prl} and the Bethe-Salpeter equation (BSE)~\cite{hank-sham80prb,stri88rnc,pusc-ambr02prb}. To understand the role of exchange-correlation effects on the electronic structure, we first compute the DFT band structure, employing both semi-local and hybrid xc functionals. Then, we obtain the quasi-particle bands by self-energy corrections. Through a detailed analysis of the dielectric function, we discuss the impact of e-h interaction on the absorption edge. We present the origin and the spatial distribution of the e-h pairs that dominate the different frequency ranges. Furthermore, we discuss the impact of e-ph coupling on the opto-electronic excitations at the onset. 
 
The theoretical absorption spectrum is compared with electron energy loss spectroscopy (EELS) performed on a single crystal. Here, we succeed using EELS at high-energy resolution as a stand-alone technique for measuring the band gap as well as obtaining the dielectric function of BaSnO$_3$. An advantage of this set-up is that transmitted electrons interact more strongly with matter than photons, transferring momentum to the orbital electrons, which enhances the measurability of indirect band gaps~\cite{egerton+11}. Moreover, it can be performed on a small crystal volume. In this way, one can select defect-free sample regions, offering the possibility to assess the properties of the perfect crystal. However, the observation of band gaps and other electronic features by EELS is typically hampered by spurious contributions from surface and relativistic losses. Also for this particular material, the relatively high refractive index at the energy region close to the band gap indicates that relativistic losses contribute to the spectrum of electrons transmitted with accelerating voltages above 60 kV. Purely experimental procedures have been proposed to generally address this issue, including low-voltage and off-axis EELS~\cite{gu+07, Stoeger+08}. Nevertheless, these procedures are material specific and can make the analysis of the spectra difficult ~\cite{kaiser+11, korneychuk+18}. Recently, inverse algorithms based on modified Kramers-Kronig analysis (KKA) are emerging as an advantageous alternative to purely experimental methods~\cite{Eljarrat+19ulm}. In this off-line approach, a self-consistent estimate of the spurious contributions to the EELS spectra is obtained and removed from the experimental data. Here, we demonstrate how this technique can be applied to reveal previously unknown valence properties and obtain a measurement of the dielectric function. With these measurement techniques, onsets of both direct and indirect transitions can be captured. To examine the possible role of indirect interband transitions on the absorption onset, the spectrum is also computed and analyzed at finite momentum transfer. Through temperature-dependent optical absorption measurements of the optical onset we determine to which extent electron-phonon coupling renormalizes the band gap, complementing the quantitative comparison between theory and experiment to give a fully consistent picture.

\section*{Results and discussion}
\subsection*{Electronic structure}
\label{sec:ES}
\begin{figure}
 \begin{center}
\includegraphics[width=.45\textwidth]{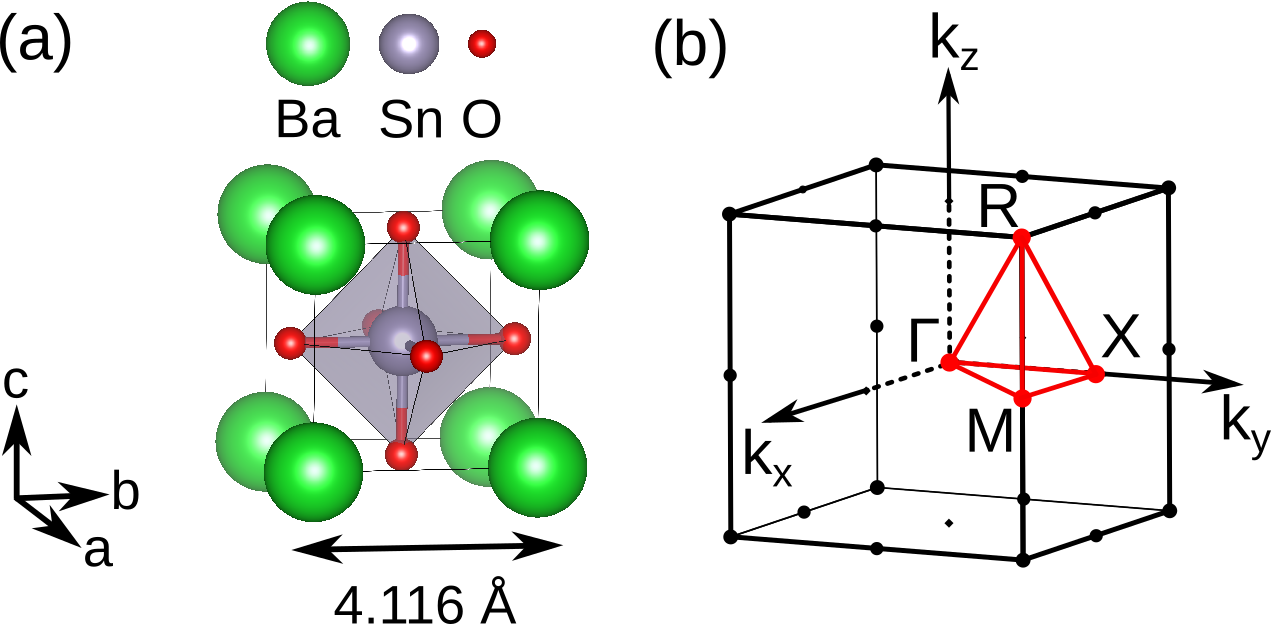}%
\caption{\textbf{Crystal structural of cubic BaSnO$_{3}$}: Primitive cell (left) and first Brillouin zone (right) with high-symmetry points and paths (red) as used in Fig.~\ref{fig:BSO-elec}.}
\label{fig:structure}
 \end{center}
\end{figure}

At room temperature, BaSnO$_{3}$ crystallizes in the cubic perovskite structure with space group Pm$\overline{3}$m as shown in Fig. \ref{fig:structure}. Its primitive unit cell containing five atoms, exhibits highly symmetric (non-tilted) SnO$_{6}$ octahedra. The lattice constant of 4.127~\AA\ obtained with the PBEsol functional is close to the experimental value of 4.116 \AA~\cite{Hkim+12ape}. As the band gap of this material is very sensitive to strain~\cite{Singh+14apl}, we adopt the latter in the following calculations such to ensure a quantitative comparison between measured and computed quantities~\cite{note-theory-data}. Considering the formal ionic charges of Ba (+2), Sn (+4), and O (-2), BaSnO$_{3}$ is formed by alternating neutral BaO and SnO$_{2}$ layers along the [100], [010], and [001] directions, making it a nonpolar material. 

\begin{figure*}
 \begin{center}
\includegraphics[width=.98\textwidth]{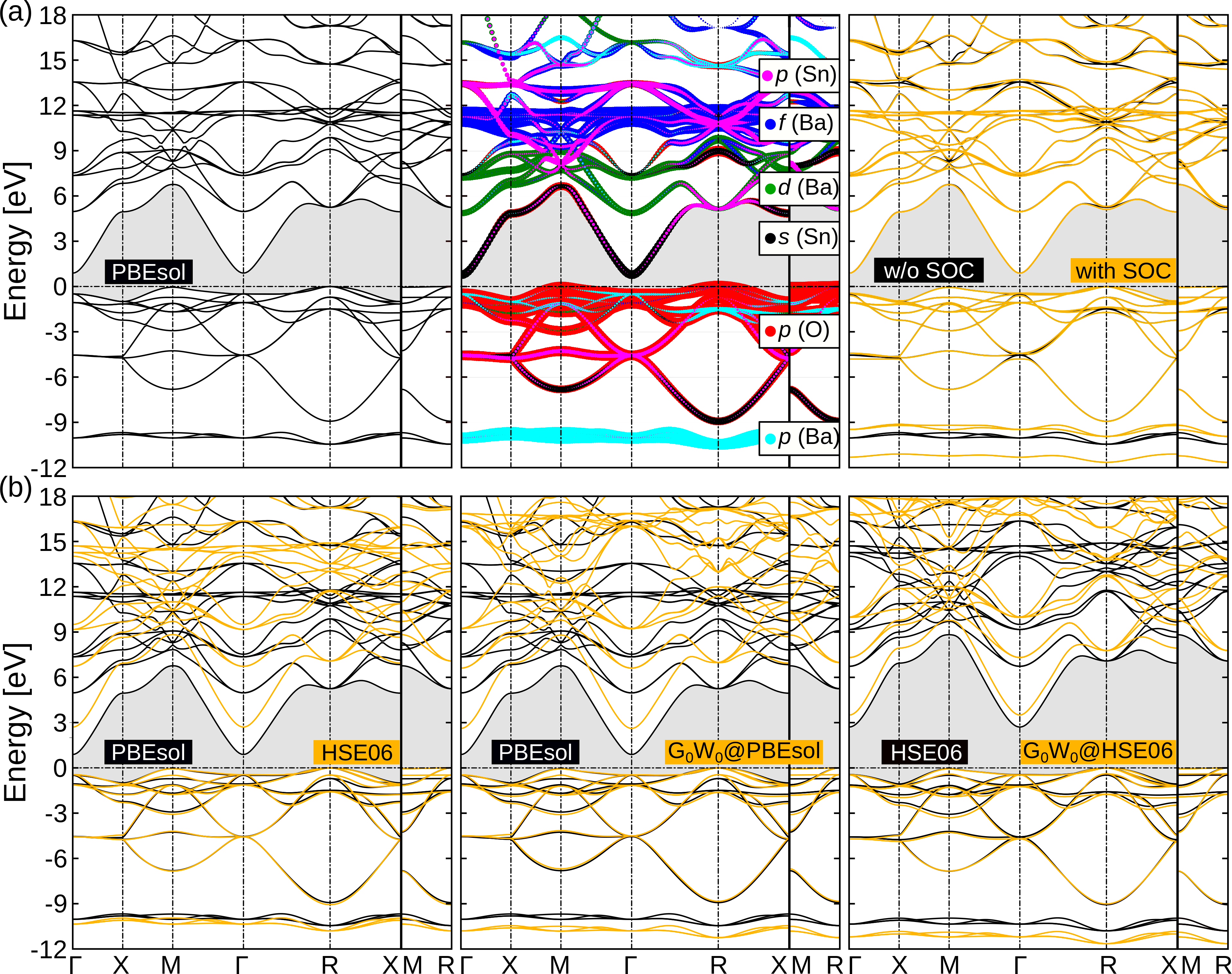}%
\caption{\textbf{Band structure of cubic BaSnO$_{3}$}. (a) Result of PBEsol, where in the middle panel the band characters are highlighted; in the right panel, the effect of spin-orbit coupling, SOC, is shown. (b) From left to right: comparison of HSE06 with PBEsol; $G_{0}W_{0}$@PBEsol; $G_{0}W_{0}$@HSE06. The valence band maximum is set to zero.}
\label{fig:BSO-elec}
 \end{center}
\end{figure*}

\begin{figure}[htbp]
 \begin{center}
\includegraphics[width=.48\textwidth]{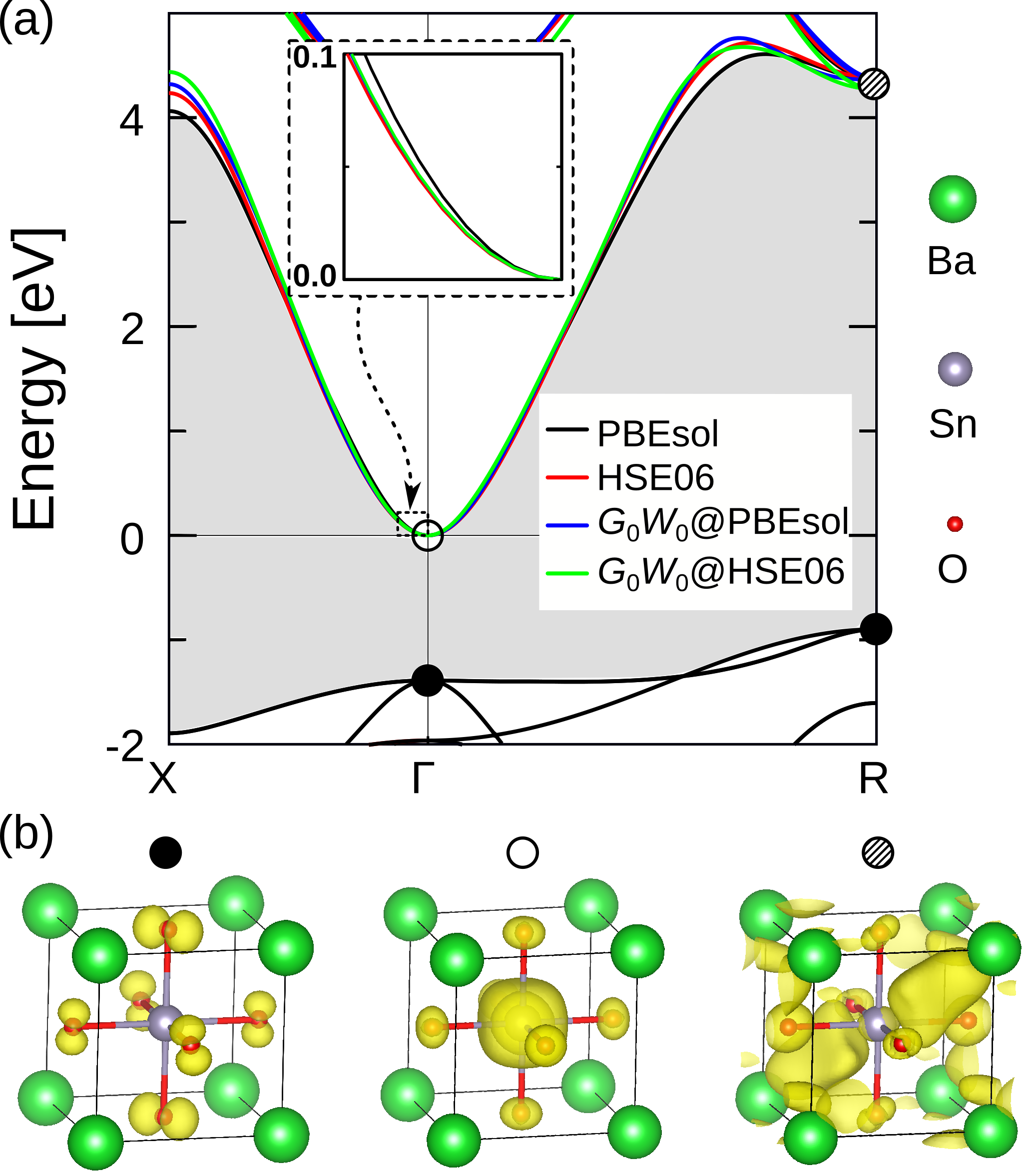}%
\caption{\textbf{Electronic structure of BaSnO$_{3}$.} (a) Conduction band obtained by PBEsol, HSE06, $G_{0}W_{0}$@PBEsol, and $G_{0}W_{0}$@HSE06. The conduction-band minimum is set to zero. The inset zooms into the region of the CBm along the $\Gamma$-X direction. (b) Square modulus of the electronic Kohn-Sham wavefunction (yellow) of the valence-band maximum at R (a similar distribution is found at $\Gamma$) as well as the CBm at $\Gamma$ and R.}
\label{fig:BSO-effective}
 \end{center}
\end{figure}

We first present the electronic band structure obtained by PBEsol and HSE06. The corresponding values of the band gaps are reported in Table~\ref{tab:gaps}. The results for PBEsol shown in the left panel of Fig.~\ref{fig:BSO-elec}(a), indicate an indirect band gap of 0.9 eV, with the valence-band maximum (VBM) and the conduction-band minimum (CBm) located at the R and the $\Gamma$ point of the Brillouin-zone, respectively. The direct gap, located at $\Gamma$, is about 1.39 eV. These values underestimate the experimental counterparts by about 70\%. As shown in the middle panels of the figure, the VBM is mainly dominated by O-\textit{p} states, while the highly dispersed conduction band around $\Gamma$ exhibits mainly Sn-\textit{s} character with some contribution from O-\textit{s} orbitals [see Fig.~\ref{fig:BSO-elec}(a) and Fig.~\ref{fig:BSO-effective}]. Ba-\textit{d} states dominate the conduction bands between 5 and 9 eV, while Ba-\textit{f}-derived bands appear above 9 eV. At the R point, the VBM and CBm are threefold degenerate, exhibiting O-\textit{p} and Ba-\textit{d} character, respectively [see also Fig.~\ref{fig:BSO-effective} (b)]. In addition, the CBm has some contribution from Sn-\textit{p} states. At the $\Gamma$ point, the VBM is also threefold degenerate, showing O-\textit{p} character. The PBEsol band structure including spin-orbit coupling is shown in the right panel of Fig.~\ref{fig:BSO-elec}(a). The effect of spin-orbit coupling is small within the energy range between -9 and 12 eV, in agreement with a previous theoretical study~\cite{soleimanpour+14pbcm}. Pronounced spin-orbit coupling appears in the valence band at about -10 eV, where the Ba-\textit{p} bands split by about 1.80 eV. Since these strongly affected bands do not enter the calculation of the optical properties, spin-orbit coupling is neglected in the following. We also note that there is no change in the size of the indirect or direct band gap due to spin-orbit coupling.

Employing the hybrid functional HSE06, the indirect (direct) band gap increases to 2.70 (3.13) eV (see Table~\ref{tab:gaps}). These values that are in agreement with previous theoretical values~\cite{Dabaghmanesh+13jpcm}, still underestimate the experimental values by about 10\%. 
Besides, we notice several changes in the band structure upon using HSE06.
As evident from the left panel of Fig.~\ref{fig:BSO-elec}(b), conduction bands can be approximated by a simple scissor shift with respect to those given by PBEsol up to 10 eV above the CBm. 
For the bands lying above, the scissor approximation is not valid due to changes in the band dispersion as well as the band order caused by the localized nature of Ba-\textit{d} and -\textit{f} states. As an example, the Ba-\textit{f}-derived bands at about 12 eV in the PBEsol band structure are shifted up by about 3 eV in HSE06.
\begin{table}[tbhp]
\centering
\caption{Band gaps, $E_{g}$, and effective electron masses, m$_{e}^{*}$/m$_{0}$, of cubic BaSnO$_{3}$, obtained by PBEsol, HSE06, $G_{0}W_{0}$@PBEsol, and $G_{0}W_{0}$@HSE06.}
\vspace{0.2cm}
 \begin{tabular}{c|c|c|ccc|cccc}
\hline
\hline
   &   & \multicolumn{2}{c}{DFT} & & \multicolumn{2}{c}{$G_{0}W_{0}$@}         \\
\cline{3-4}
\cline{6-7}
    &    &PBEsol&HSE06       & & PBEsol& HSE06 \\
\hline 
\multirow{2}*{E$_{g}$[eV]} &R-$\Gamma$ &  0.90 & 2.70  &  & 2.62 & 3.50 \\
 \cline{2-7}
              & $\Gamma$-$\Gamma$ &  1.39 &  3.13  &  & 3.08 & 3.96 \\
 \cline{1-7} 
m$_{e}^{*}$/m$_{0}$     &$\Gamma$-X &0.17&  0.21  &  & 0.20 & 0.20\\
\hline
\hline
\end{tabular}
\label{tab:gaps}
\end{table}

To obtain the quasi-particle (QP) band structure, we apply a self-energy correction (Eq.~\ref{eq:self}) in the $G_{0}W_{0}$ approximation. To this end, we consider both the PBEsol and the HSE06 band structure as a starting point, and the results are depicted in the middle and right panel of Fig.~\ref{fig:BSO-elec}(b), respectively. As shown in Table~\ref{tab:gaps}, the QP correction increases the indirect (direct) gap by approximately 1.7 eV, resulting in 2.62 (3.08) eV. Such a large magnitude of the self-energy correction reflects the strong electron-electron correlation effects in this material. Therefore, a pronounced starting-point dependence is expected. The QP gaps obtained starting from PBEsol are close to those of plain HSE06. $G_{0}W_{0}$ on top of the latter ($G_{0}W_{0}@$HSE06) leads to a further increase by about 0.8 eV, giving values of 3.50 (3.96) eV for the indirect (direct) gap. This method has turned out most reliable for computing the electronic properties of a wide range of oxides, including TCOs~\cite{Bechstedt+17jmr}. However, as expected, these values are larger than most experimental gaps as neither ZPV nor temperature-effects are considered here. We will get back to this point in Sections~Optical properties. 

Looking at the entire band structures, the QP correction to both PBEsol and HSE06 varies from band to band and depends on the \textbf{k}-point, specifically, within the bands where the localized Ba-\textit{d} and Ba-\textit{f} states appear. The bands derived from the latter (not shown) are located about 14 eV above the CBm in both $G_{0}W_{0}@$PBEsol and $G_{0}W_{0}$@HSE06 [see Fig.~\ref{fig:BSO-elec}(b)]. Overall, a simple scissor shift of the DFT conduction bands is sufficient to capture the features of the unoccupied bands up to 10 eV above the CBm. However, for computing optical spectra in a larger energy window ($>$10 eV) the QP band structure is needed~\cite{yan+18jvsta}.
Due to the computational complexity, we use the QP band structure obtained by $G_{0}W_{0}@$PBEsol and a scissors operator of 0.88 eV to recover the $G_{0}W_{0}$@HSE06 band gap.

Finally, we focus on the effective electron masses for which, previous calculations have shown that it is sensitive to the xc-functional~\cite{dongmin+14apl,kim+13jssc}. Our results are summarized in Table~\ref{tab:gaps}. For visual inspection, we plot in Fig.~\ref{fig:BSO-effective}(a) all band structures computed in this work in the vicinity of the band gap, setting the CBms to zero. Although there exists a dependency on the applied method, the differences stay within approximately 15\%. While PBEsol and HSE06 calculations give about 0.21 and 0.17 m$_{\mathrm{0}}$, respectively, along the $\Gamma$-X direction and the same values are found also along the $\Gamma$-M and $\Gamma$-R directions, we obtain 0.20 m$_{\mathrm{0}}$ with both $G_{0}W_{0}@$PBEsol and $G_{0}W_{0}$@HSE06. Comparing with previous calculations, our results are close to those obtained by HSE06~\cite{krish+17prb,Krish+16apl,Dabaghmanesh+13jpcm}.
 
\subsection*{Optical properties}
We proceed with our analysis by investigating the optical properties of BaSnO$_{3}$. Previous investigations reported the material to be transparent in the visible region and absorbing light from the near-UV range above 3 eV, with relatively low intensity, followed by intense absorption starting from 7 eV~\cite{Hkim+12prb,monserrat+18prb,yan+18jvsta,Stanislavchuk+12jap,moreira+12jap}. Difficulties to analyze the shape of the experimental absorption edge arise mainly due to the fact that the absorption onset originates from weak phonon-assisted transitions across the indirect band gap ~\cite{monserrat+18prb,Hkim+12prb,galazka+16jpcm}. Further complications arise from the presence of defects and impurities~\cite{Stanislavchuk+12jap}. EELS experiments in the scanning transmission electron microscope (STEM-EELS) turns out to be an advantageous experimental technique to address these difficulties. One reason already mentioned is that momentum exchange increases the relevance of indirect transitions. Additionally, when performed in the STEM-EELS mode, properties from relatively small volumes of material can be measured. This makes EELS desirable when larger defect-free samples are difficult to obtain. To get insight into the characteristics of the absorption onset, we present and analyze in the following computed and measured absorption spectra.

\subsubsection*{BSE calculations}
\label{sec:optics}

\begin{figure}
 \begin{center}
\includegraphics[width=.47\textwidth]{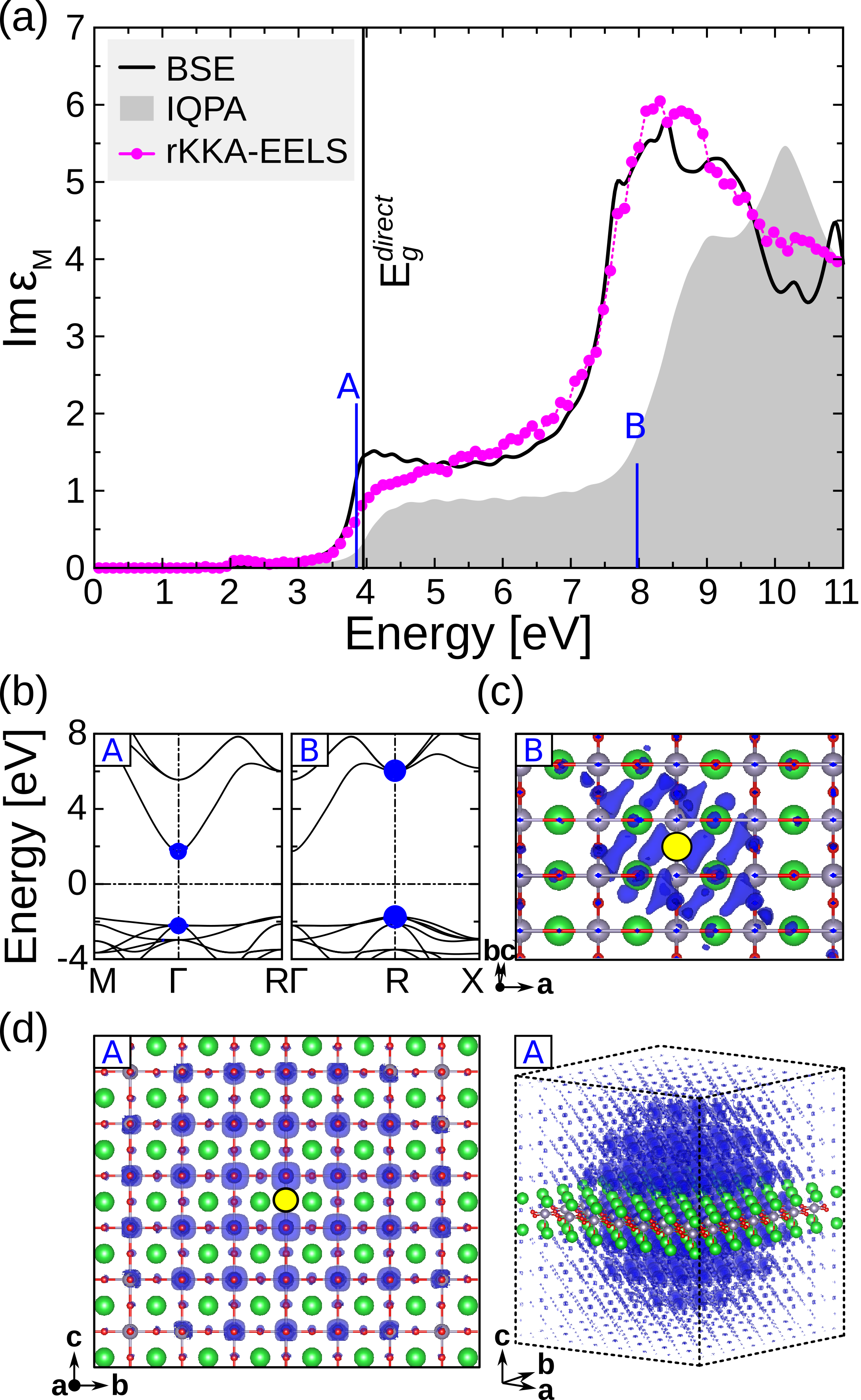}%
\caption{\textbf{Optical absorption of cubic BaSnO$_{3}$}. (a) Imaginary part of the macroscopic dielectric function obtained from the Bethe-Salpeter equation (BSE) which includes excitonic effects (solid line). The independent-quasiparticle spectrum (IQPA), is shown for comparison. A Lorentzian broadening of 0.15~eV is applied to both spectra to mimic excitation lifetimes. The experimental result (magenta circles) is obtained by applying relativistic Kramers-Kronig analysis (rKKA) to electron energy-loss spectroscopy (EELS) data. (b) Band contributions to excitations (A) and (B) marked in the BSE spectrum. (c, d) Spatial distribution of the electron-hole wavefunction for excitations (A) and (B), with the hole fixed on the O atom (yellow spots). In (d), we also show the three dimensional electron distribution of excitation A, where only a few atomic layers of the BaSnO$_{3}$ structure are shown to ease the visualization.}
\label{fig:BSO-optics}
 \end{center}
\end{figure}

In Fig.~\ref{fig:BSO-optics}(a), we depict the optical spectra of BaSnO$_{3}$ computed by the BSE. In the displayed energy range, we identify two main regions, distinguished by their intensity. Between 3.86 eV and 7 eV (region I), the absorption is relatively low. Above that (region II), the intensity increases, and a pronounced peak centered at about 8 eV is formed. The electron-hole interaction vastly redshifts the entire spectrum as observed from comparison with the independent particle spectrum. The absorption starts at 3.86 eV (optical gap), which is 100 meV below the direct gap (3.96 eV) where the independent particle onset sets in. Region I is governed by transitions between the O-\textit{p}-derived valence bands and the Sn-\textit{s}-derived conduction band. Bound excitons give rise to high spectral intensity in the region around the direct band gap. The lowest bound state, exciton A, arises from transitions around the VBM and the CBm at the $\Gamma$ point of the Brillouin-zone as highlighted in Fig.~\ref{fig:BSO-optics}(b). The corresponding electron-hole wavefunction is shown in Fig.~\ref{fig:BSO-optics}(d) for a fixed hole position at the oxygen site. The electron is delocalized around the hole with a 3D extension larger than 20~\AA, mainly spreading over the Sn-\textit{s} and the O-\textit{s} orbitals. This delocalized character reflects the rather low exciton binding energy of about 100 meV which is comparable to values in other oxides, {\it e.g.}, in ZnO~\cite{Gori+10prb} and LaInO$_{3}$~\cite{Aggoune+LIO}.

Starting from 7 eV (region II), the spectral weight increases steeply, where the peak at around 8 eV stems from intensive transitions between O-\textit{p} and Ba-\textit{d} states. From the comparison with the corresponding peak in the independent particle spectrum, one can notice a significant red-shift by $\sim$2 eV, induced by electron-hole attraction [Fig.~\ref{fig:BSO-optics}(a)] while local-field effects are negligible. Exemplary, we consider excitation B that contributes to the peak. As evident from Fig.~\ref{fig:BSO-optics}(b), B is built of transitions between the VBM and the CBm at the R point [see also Fig.~\ref{fig:BSO-effective}(b)]. The corresponding excitonic wavefunction is characterized by a rather localized electronic distribution (2-3 unit cells) surrounding the hole whose position is fixed on an O atom for visualization [Fig.~\ref{fig:BSO-optics}(c)]. The electrons spread mainly over Ba-\textit{d} and to less extent over Sn-\textit{s} orbitals.

\subsubsection*{EELS measurements}
\label{Sec:eels}

In Fig.~\ref{fig:BSO-optics}(a), we also display the imaginary part of the dielectric function obtained by applying relativistic Kramers-Kronig analysis (rKKA) to the EELS measurement. One can see excellent agreement with the BSE result in terms of peak positions, magnitude, and overall shape. The real part of the obtained dielectric function (not shown) indicates that the threshold for Cherenkov radiation is considerably surpassed only for the energy region above the band gap, {\it i.e.}, $\sim 5-10$ eV. In fact, our result indicates that the impact of relativistic losses for the band-gap region is small, considering the experimental parameters and estimated thickness of this sample. For more details and comparison between theory and experiment of both real part and imaginary part at higher energies, we refer to the Supplementary Figure 3. Focusing on the location of the absorption onset, we note that even after plasma cleaning (see Method section), there is still some weak signal seen at about 2 eV [Fig.~\ref{fig:BSO-optics}(a)]. This might be mainly attributed to a remaining contamination at the BaSnO$_{3}$ surface, but it apparently does not affect the spectrum at higher energies. Therefore, as depicted in Fig.~\ref{fig:BSO-q}(a), we analyze the onset of the EELS spectrum in order to understand the characteristic of the absorption edge. 

At first glance, we recognize two regimes, consisting of a region between 3 and $\sim$3.5 eV where the signal only increases weakly, followed by a drastically increasing intensity above 3.5 eV. While the latter stems from strong direct transitions, we attribute the former to originate from indirect transitions. This characteristic is inline with recent experimental observations from optical absorption~\cite{galazka+16jpcm}, as well as with angle-resolved photoemission spectroscopy~\cite{beom+17cap}. This finding is also supported by calculations on the independent-particle level~\cite{monserrat+18prb}. 

 \renewcommand{\arraystretch}{1.3}
\begin{table}[h]
\centering
\caption{Absorption onset, E$^{onset}$, given by the energy of the first indirect transition, and optical gap, E$_g^{opt}$, determined by the first direct excitation, as obtained by the EELS and optical absorption measurements~\cite{galazka+16jpcm}.}
\vspace{0.2cm}
 \begin{tabular}{c|c|c|}
Experiment &  E$^{onset}$ [eV] & E$_g^{opt}$ [eV] \\
\hline 
EELS at 300 K & 2.97$\pm$0.04 & 3.57$\pm$0.04 \\
Optical absorption at 5 K &  3.17$\pm$0.04  &-\\
Optical absorption at 300 K &  2.99$\pm$0.04  &-\\
\end{tabular}
\label{tab:exp}
\end{table}
To determine the absorption from the EELS spectra, we employ the model introduced in Section~Method, restricting the fit to the energy range between 2.5 to 4.5 eV (see Supplementary Methods for more details). The model parameters are adjusted using bounded optimization with the Levenberg-Marquart algorithm followed by dual annealing as implemented in the Python package HyperSpy~\cite{hyperspy}. Our results indicate that the onset of the indirect and direct transitions are at 2.97$\pm$0.04 eV and 3.57$\pm$0.04 eV, respectively. The difference between the two onsets of about 0.60$\pm$0.06 eV agrees well with the calculated $G_{0}W_{0}$@HSE06 counterpart of $\sim$0.45 eV (Table~\ref{tab:gaps}). As shown in Table~\ref{tab:exp}, excellent agreement is found with the optical measurement of Ref.~\onlinecite{galazka+16jpcm}. Overall, the measured values are also in agreement with most reported experimental values that are about $\sim$3.0~\cite{yan+18jvsta,rylan+20jcg,Kang+18apl} and 3.5 eV~\cite{Stanislavchuk+12jap,Kang+18apl,chambers+16apl,beom+17cap} for the onset of the indirect and direct transitions, respectively. Note that both values are not fundamental band gaps as the spectra include excitonic effects. A detailed comparison between experiment and theory will be given in the last part before conclusions.

We also estimate the effective electron mass from the dielectric function in the framework of the free-electron approximation~\cite{Gass+06prb},
\begin{equation}
m^{*}=\frac{n_{eff}}{\varepsilon_{0}\varepsilon_{\infty}}\left(\frac{e\hbar}{E_{pl}}\right)^{2},
\end{equation}
where $n_{eff}$ is the effective valence electron density, $\varepsilon_{\infty}$ is the high-frequency dielectric constant, and $E_{pl}$ is the plasmon energy of the material. The latter is retrieved as the zero-crossing of the real part from the negative half-plane (see Supplementary Figure 3). The obtained value of E$_{pl}\sim$25.2 eV is in excellent agreement with the free-electron prediction for this material using the experimental lattice parameter of 4.116 \AA. The high-frequency dielectric constant, obtained from the rKKA analysis is $\varepsilon_{\infty}\sim$4.2, leading to an estimate for the effective electron mass of $m^{*}\sim$0.16 $m_{0}$. This value is very close to the PBEsol result of $\sim0.17~m_{0}$ and slightly lower but still comparable to the $G_{0}W_{0}$@HSE06 value of 0.20 $m_{0}$. The deviation could be due to the usage of the simple free-electron framework. Overall, these values confirm experimental works reporting a low effective mass of $\sim$0.19 m$_{\mathrm{0}}$~\cite{Niedermeier+17aps,james+16apl}, which suggests a high electron mobility in BaSnO$_{3}$.\\

\subsubsection*{Optical absorption}
\label{Sec:opt-abs}

\begin{figure}[h!]
\includegraphics[width=.48\textwidth]{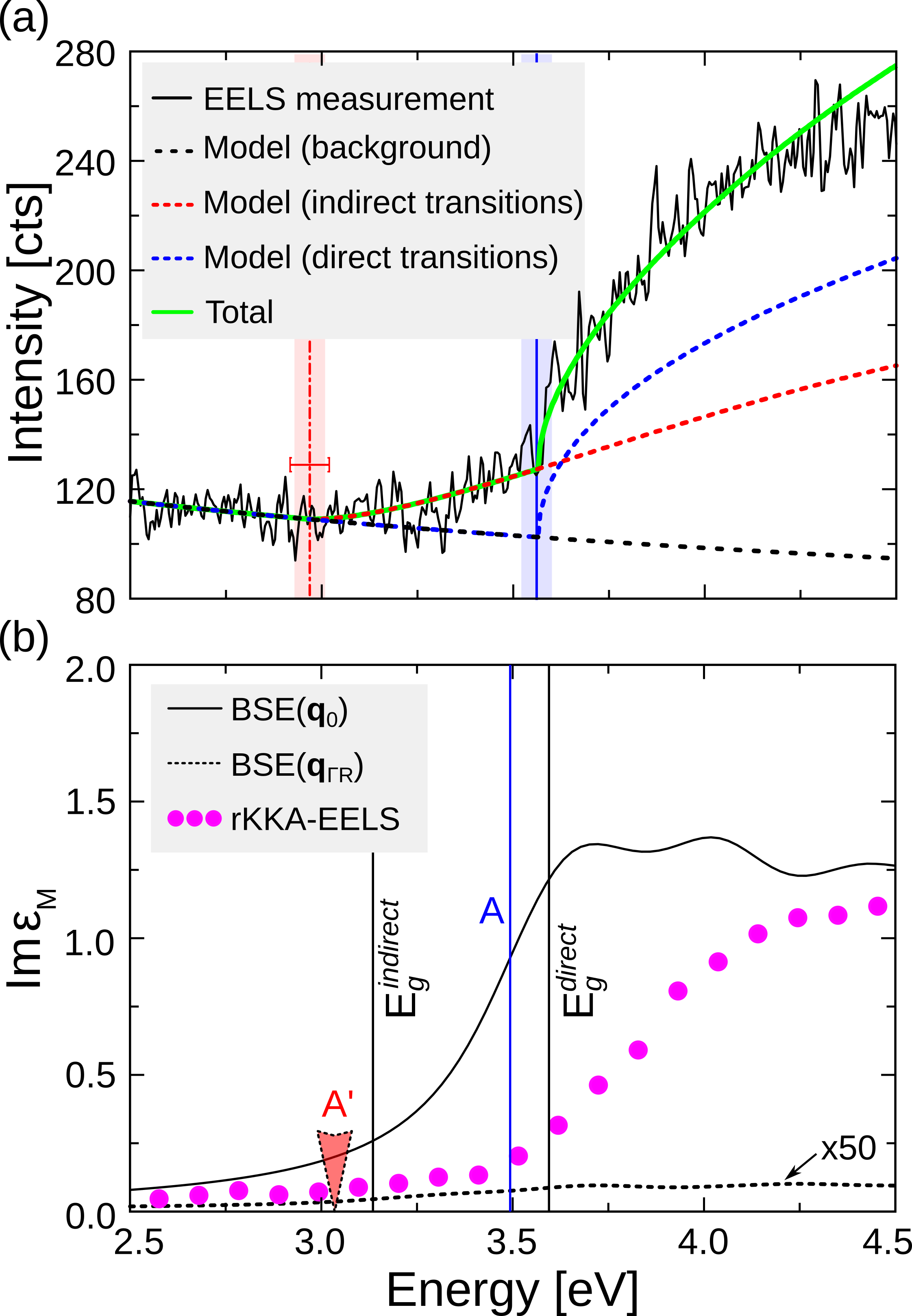}
\caption{\textbf{Comparison between theory and experiment.} (a) Electron energy-loss spectroscopy (EELS) measurement (black solid line) of a BaSnO$_{3}$ single crystal in the energy range around the absorption onset. The fits obtained by a model assuming either indirect or direct transitions (see Method section) are shown by the red and blue dashed lines, respectively. The dashed black line refers to the background model while the green solid line shows the sum of the three models. The vertical blue line indicates the onset of direct transitions and the dashed red line that of indirect transitions; the shaded areas represents their error bar. (b) Dielectric function in the energy range of the absorption onset given by Bethe-Salpeter equation (BSE, black solid line) and determined from EELS measurement (magenta circles). The intensity of the BSE spectrum for momentum transfer \textbf{q}=\textbf{q}$_{(\Gamma R)}$=(1/2,1/2,1/2), shown by the dashed line, is scaled by a factor of 50 for better visibility. A Lorentzian broadening of 0.2 eV is applied. The blue line indicates the first direct excitation (optical gap) obtained for \textbf{q}$_{0}$=(0,0,0) (A), the red arrow refers to the first indirect transition for \textbf{q}=\textbf{q}$_{\Gamma R}$ (A'). Both BSE spectra are corrected by a value of -0.367 eV to account for band-gap renormalization due to zero-point vibrations and temperature.}
\label{fig:BSO-q}
\end{figure}
For a quantitative evaluation of band gaps and absorption onsets, vibrational effects must be considered. Previous finite-temperature calculations~\cite{monserrat+18prb} indicate that the onset of the indirect transitions at 300 K is about 0.19 eV below that at 0 K. The experimental counterpart was estimated to be about 0.18 eV (Table~\ref{tab:exp})~\cite{galazka+16jpcm}. In both cases, the role of ZPV was not known. To remedy this situation, we perform optical absorption measurements below room temperature. The resulting dependence of the energy onset is shown in Fig.~\ref{fig:BSO-ZPV}. More information related to the optical absorption spectra, crystal growth, and sample preparation can be found in Ref.~\onlinecite{galazka+16jpcm}.

The measured data are fitted using a single-oscillator model~\cite{Donnell+91apl}, expressed by
\begin{equation}
E^{\mathrm{onset}}(T)=E^{\mathrm{onset}}(0)-S\langle \hbar\omega \rangle \left[\mathrm{coth}\left( \frac{\langle \hbar\omega \rangle}{2k_BT}\right)-1\right],  
\end{equation}
where the parameter $E^{\mathrm{onset}}(0)$ is the energy onset at zero temperature, $S$ a measure of the electron-phonon coupling strength, and $\langle\hbar\omega\rangle$ an average phonon energy~\cite{Donnell+91apl,irmscher+14pssa,Aggoune+LIO}. Considering the whole temperature range, the best-fit parameters $E^{\mathrm{onset}}(0)$= 3.17~eV, $S$= 6.5, and $\hbar\omega$= 28~meV are obtained. Extrapolating to 1200 K, the energy is by about 1 eV lower than at room temperature, reflecting the strong impact of electron-phonon coupling. Similar values have also been reported for other TCOs such as In$_{2}$O$_{3}$ and SrTiO$_{3}$~\cite{kok+15pssa,irmscher+14pssa}. The high-temperature part of the curve can be fitted with a linear regression, {\it i.e.}, $E^{\mathrm{onset}}(0)-2Sk_BT+ S\langle \hbar\omega \rangle$~\cite{Donnell+91apl}. The ZPV contribution is, thus, given by $S\langle \hbar\omega \rangle$ which amounts to 0.182~eV. Going from 0 and 300 K, the onset decreases by about 0.367 eV, where 0.182 eV is assigned to ZPV and 0.185 eV to temperature effects including lattice expansion. These values are comparable to those reported for other TCOs~\cite{irmscher+14pssa,Aggoune+LIO}. These values will be used as corrections to enhance the theoretical spectra as shown in Fig.~\ref{fig:BSO-q}(b).

\subsubsection*{Comparison between theory and experiment}
\label{Sec:theoy-vs-exp}

\begin{figure}
 \begin{center}
\includegraphics[width=.49\textwidth]{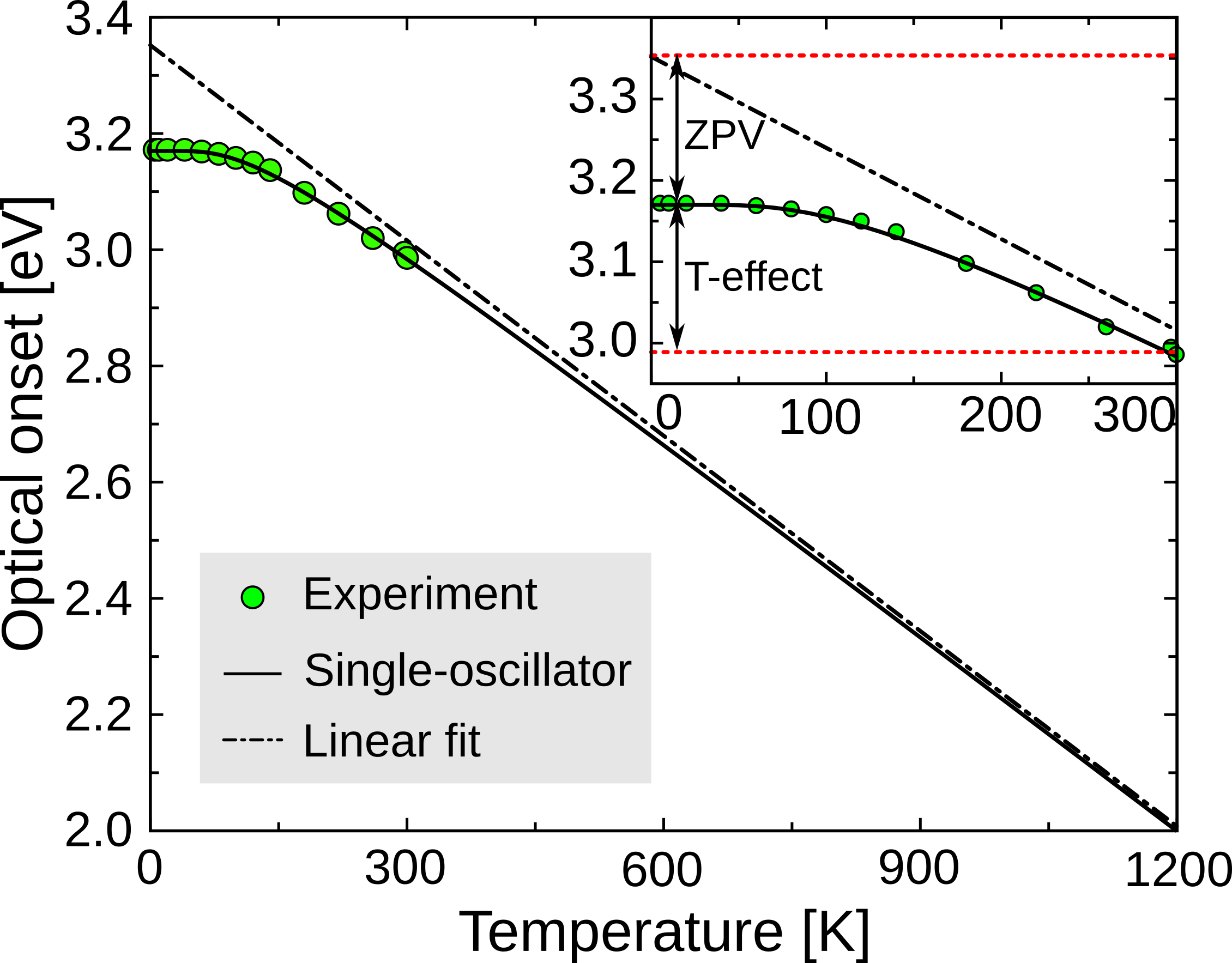}%
\caption{\textbf{Temperature dependence of the absorption onset obtained by optical-absorption measurements}. The circles represent the measured data; the corresponding solid line is the best fit based on a single-oscillator model. The dashed line shows a linear fit for high temperatures. The inset zooms into the low-temperature region to distinguish between gap-renormalization due to temperature (T-effect) and zero-point vibrations (ZPV).}
\label{fig:BSO-ZPV}
 \end{center}
\end{figure}

The BSE spectrum shown in Fig.~\ref{fig:BSO-optics}(a) is formed only by vertical excitations. To describe transitions across the indirect gap, {\it i.e.}, from R to $\Gamma$, we solve the BSE for momentum transfer \textbf{q$_{\Gamma R}$}=(1/2,1/2,1/2). Both spectra are depicted in Fig.~\ref{fig:BSO-q} (b), where the correction to account for vibrational effects, estimated above, is included in the scissors shift (see Method section).

We note that the latter spectrum has almost zero intensity at the onset, compared to that of \textbf{q}$_{0}$. This is attributed to the use of centrosymmetric BaSnO$_{3}$ structure, \textit{i.e.}, we do not consider atomic displacements due to phonons. As such, transitions between the threefold degenerate VBM at the R point and the CBm at the $\Gamma$ point belong to the T$_{2u}$ and A$_{1g}$ representations of point group O$_{h}$, respectively~\footnote{This analysis was carried out by using the $\mathrm{Irvsp}$ tool~\cite{irvsp} of the VASP code.}, are dipole forbidden. However, mediated by vibrations, such transitions become optically allowed, as supported by calculations on the independent-particle level~\cite{monserrat+18prb}. Therefore, we attribute the measured weak absorption across the indirect band gap here [Fig.~\ref{fig:BSO-q}(a)] and in Ref.~\cite{galazka+16jpcm} to the fact that phonons lower the symmetry of the here simulated ideal crystal. Such effect has previously been observed in other materials~\cite{Morris+18prb,Olocsson+19jpcc,blaha+18prb}. Importantly, our BSE calculation at finite momentum allows us to access the location of the first indirect transition as well as the amount of the excitonic effects. We note that the differences between the rKKA-EELS and the BSE (q=0) spectrum in Fig.~\ref{fig:BSO-q}(b) is mainly due to ignoring temperature in the latter which has been reported to reduce the intensity at the onset~\cite{monserrat+18prb}.
 
\renewcommand{\arraystretch}{1.3}
\begin{table}[h]
\centering
\caption{Indirect QP gap, E$_g^{QP}$, as given by $G_{0}W_{0}$@HSE06 at 0K and optical gap, E$_g^{opt}$, obtained by the BSE [in eV]. In addition to the energy of the first direct excitation (A), defining the optical gap, also the energy of the first indirect excitation (E$^{A'}$) is provided. The corresponding values including corrections (marked with "-corr") to account for ZPV and temperature effects, estimated experimentally, are depicted as well. For comparison, the related experimental gap values (Table II) are given in brackets.}

\vspace{0.2cm}
 \begin{tabular}{c|cc|cc|cc}
Theory &             E$_g^{QP}$  & E$_g^{QP-corr}$ &E$^{A'}$ &  E$^{A'-corr}$  & E$_g^{opt}$ &  E$_g^{opt-corr}$  \\
\hline
0 K                 & 3.50   &  3.32 & 3.40  & 3.22 (3.17) &  3.86&  3.68\\
300 K  &     & 3.13 &&  3.03($\sim$3.0) &&     3.50(3.57)  \\

\end{tabular}
\label{tab:zpv-qp}
\end{table}
In Table~\ref{tab:zpv-qp}, we report the indirect QP gap, as given by $G_{0}W_{0}$@HSE06 at 0K, and the optical gap obtained by the BSE. The latter is given by the first direct excitation (A) [Fig.~\ref{fig:BSO-q}(b)]. In addition, we also provide the energy of the first indirect excitation (A'). The corrected values, accounting for ZPV and temperature effects shown in Fig.~\ref{fig:BSO-ZPV}, are also shown. At room temperature, the fundamental QP gap is 3.13 eV. We obtain the absorption onset (excitation A') and the optical gap (excitation A) at room temperature to be 3.03 and 3.5 eV, respectively. The former is in excellent agreement with both EELS (2.97$\pm$0.04 eV) and optical absorption (2.99$\pm$ 0.04) counterparts (Fig.~\ref{fig:BSO-q} eV). The latter value is also in a very good agreement with the optical gap obtained by the EELS measurement (3.57$\pm$0.04 eV) (see Fig.~\ref{fig:BSO-q}). 

Finally, the binding energy of excitation A' amounts to 100 meV, the same value as that of the direct excitation A [see Fig~\ref{fig:BSO-q}(b)]. We note that excitonic effect in the EELS measurement may be smeared out at room temperature. Anyway, the predicted, relatively small computed value is within the experimental error bar. Considering the measured absorption onset at 3.0 eV, we can deduce the fundamental gap by adding the excitonic binding energy. Doing so, we arrive at a value of $E_{g}^{exp}$=3.1 eV at room temperature.

\section*{Conclusions}
We have presented a detailed analysis of the electronic and optical properties of the cubic perovskite BaSnO$_{3}$ from a first-principles, many-body study and experiment. Applying a self-energy correction to the bands obtained by the hybrid functional HSE06 ($G_{0}W_{0}$@HSE06), we predict an indirect (direct) QP gap of 3.50 (3.96) eV and an effective electron mass of about 0.2 m$_{\mathrm{0}}$. Spin-orbit coupling has a minor effect on the band edges. The optical gap obtained by BSE calculations is 3.86 eV (not considering vibrational effects), reflecting an exciton binding energy of 100 meV of the lowest-energy excitation. The latter has a delocalized character, distributed over Sn-\textit{s} and O-\textit{s} orbitals. Above 7 eV, Ba-\textit{d} states are involved in the transitions and give rise to more localized excitons with high binding energy.

Temperature-dependent optical absorption measurements show the optical onset to decrease by about 0.185 eV going from 5 to 300 K. Below that, ZPV effects are estimated to amount to a further reduction of 0.182 eV. Including the latter corrections, the BSE spectrum shows an excellent agreement with the experimental counterpart obtained from the EELS measurement, in terms of energy, intensity, and shape. The latter captures an additional weak absorption feature below the strong direct transitions. It is assigned to indirect transitions driven by phonons, as reported previously in a theoretical work~\cite{monserrat+18prb}. The onset of direct transitions (optical gap) is observed at 3.57$\pm$0.04 eV, in good agreement with the theoretical value of 3.50 eV. The calculated spectrum at finite momentum transfer shows that the lowest excitation across the indirect gap has also a binding energy of about 100 meV. It sets in at about 3.03~eV, in excellent agreement with its counterparts obtained by EELS (2.97$\pm$0.04 eV) and by optical absorption (2.99$\pm$ 0.04 eV). As such indirect excitations are dipole forbidden and thus should not be observed in the latter experiment for a perfect crystal, our findings support the theoretical observation \cite{monserrat+18prb} of the weak absorption across the indirect gap to be related to phonon-induced symmetry lowering.

Overall, we have demonstrated that the combination of state-of-the-art techniques in both theory and experiment and careful analysis allow for a quantitative characterization of excitations in crystalline materials as exemplified by BaSnO$_{3}$.\\

\section*{Methods}
\subsection*{First-principles calculations}

Ground-state (GS) properties are calculated using DFT~\cite{hohe-kohn64pr,kohn-sham65pr} with the generalized gradient approximation in the PBEsol parameterization~\cite{PBEsol+08prl} for the exchange-correlation functional. Also the hybrid functional HSE06~\cite{HSE+06jcp} with 25\% of Hartree-Fock exchange is employed to compute the electronic properties. 
 
Quasi-particle (QP) energies are obtained by the $G_{0}W_{0}$ approximation \cite{hedi65pr,hybe-loui85prl} by solving the QP equation
\begin{equation}
\epsilon_{i}^{QP}=\epsilon_{i}
^{KS}+\langle\phi_{i}^{KS}\vert\Sigma(\epsilon_{i}^{QP})
-v_{XC}^{KS}\vert\phi_{i}^{KS}\rangle,
\label{eq:self}
\end{equation}
where $\Sigma$ is the non-local and energy dependent electronic self-energy, $\epsilon_{i}^{KS}$ and $\phi_{i}^{KS}$ are the Kohn-Sham energies and wave-functions, respectively. Band structure and effective masses are computed by making use of the Wannier interpolation~\cite{seb+20prb}.

The optical spectra are obtained by solving the Bethe-Salpeter equation (BSE), the equation of motion of the two-particle Green function~\cite{hank-sham80prb,stri88rnc,pusc-ambr02prb}. This problem can be mapped onto the secular equation 
\begin{equation}
\sum_{v'c'\mathbf{k'}} H^{BSE}_{vc\mathbf{k},v'c'\mathbf{k'}}A^{\lambda}_{v'c'\mathbf{k'}} = E^{\lambda}A^{\lambda}_{vc\mathbf{k},}
\label{eq:ham}
\end{equation}
where $v$, $c$ and \textbf{k} indicate valence bands, conduction bands, and \textbf{k}-points in the reciprocal space, respectively. The effective Hamiltonian consists of three terms, $H^{BSE} = H^{diag} + H^{dir} + 2H^{x}$. The first term, $H^{diag}$, accounts for vertical transitions between QP energies and, when considered alone, corresponds to the independent quasiparticle approximation (IQPA). The other two terms incorporate the screened Coulomb interaction ($H^{dir}$) and the bare electron-hole exchange ($H^{x}$). The factor 2 in front of the latter accounts for the spin multiplicity in non-spin-polarized systems. The eigenvalues of Eq.~\eqref{eq:ham}, $E^{\lambda}$, are the excitation energies. The corresponding eigenvectors, $A^{\lambda}_{vc\mathbf{k}}$, provide information about the composition of the $\lambda$-th excitation and act as weighting factors in the transition coefficients
\begin{equation}
\mathbf{t}_{\lambda} = \sum_{vc\mathbf{k}} A^{\lambda}_{v c \mathbf{k}} \frac{\langle v \mathbf{k} \vert \widehat{\mathbf{p}} \vert c \mathbf{k} \rangle}{\epsilon_{c \mathbf{k}}\ -\ \epsilon_{v \mathbf{k}}},
\label{eq:osci} 
\end{equation}
that determine the oscillator strength in the imaginary part of the macroscopic dielectric function,
\begin{equation}
\mathrm{Im}\varepsilon_M~=~\dfrac{8\pi^2}{\Omega} \sum_{\lambda} |\mathbf{t}_{\lambda}|^2 \delta(\omega - E^{\lambda}),
\label{eq:abs}
\end{equation}
where $\Omega$ is the unit cell volume. 

All calculations are performed using \texttt{exciting}~\cite{gula+14jpcm,nabo+16prb,vor+es19}, an all-electron full-potential code, implementing the family of linearized augmented planewave plus local orbitals ((L)APW+LOs) methods. The atomic radii are chosen 2.2 bohr for Ba, 2.0 bohr for Sn, and 1.6 bohr for O. The local part of the LAPW basis includes three \textit{s}, \textit{p}, and \textit{d} local orbitals for Ba and Sn, and three \textit{s} and \textit{p} local orbitals for O. The Sn 4\textit{s} and 4\textit{p} orbitals are treated as semicore states. The LAPW cutoff G$_{max}$ is set to 5.0 bohr$^{-1}$ what guarantees adequate convergence of the groundstate properties. Spin-orbit coupling is included by use of a second-variational procedure. Brillouin-zone integrations are carried out on an 8$\times$8$\times$8 $\textbf{k}$-grid for all total-energy and electronic-structure calculations. Lattice constants and internal coordinates are optimized until the residual force on each atom is less than 0.005 eV/\AA. In the HSE06 calculations, a basis-set cutoff G$_{max}$=4.375 bohr$^{-1}$ and 250 unoccupied states are used. The $G_{0}W_{0}$ quasi-particle band structures is computed using as starting point either PBEsol or HSE06 eigenstates. Here, the same computational parameters are chosen as for HSE06, reaching a numerical precision of about 30 meV for the band gap (see Supplementary Figure 1). The frequency-dependent screened Coulomb potential is treated in the random-phase approximation and computed on a grid of 16 imaginary frequencies. Pad\'{e} approximation is employed for evaluating the self-energy on a real frequency mesh to solve the quasi-particle equation Eq.~\ref{eq:self}. 

For the solution of the BSE~\cite{vor+es19} on top of the QP band structure, an LAPW cutoff G$_{max}$ of 4.375 bohr$^{-1}$ is employed. The screened Coulomb potential is computed using 100 empty bands. In the construction and diagonalization of the BSE Hamiltonian, 9 occupied and 7 unoccupied bands are included, and a 14$\times$14$\times$14 shifted $\textbf{k}$-point mesh is adopted. Owing to the high dispersion of the conduction band around its minimum at the $\Gamma$ point, the \textbf{k}-point sampling is interpolated onto a 28$\times$28$\times$28 mesh, using a double grid technique~\cite{gille+13prb}. These parameters ensure well converged spectra within an energy window up to 10 eV, and a precision of the binding energy of the lowest-energy exciton within 20~meV (see Supplementary Figure 2). For the \textbf{q}-dependent absorption spectra~\cite{vor+es19}, we solve the BSE for a momentum transfer of \textbf{q}$_{(\Gamma R)}$ between $\Gamma$ and R using the centorsymetric structure ({\it i.e.}, without considering atomic displacement due to phonons). Since we focus on the absorption onset, it is sufficient to use the PBEsol band structure and apply a scissor operator. In this case, three occupied and two unoccupied bands are included for the construction and diagonalization of the BSE Hamiltonian, and a 20$\times$20$\times$20 shifted $\textbf{k}$-point mesh is adopted. This parameter choice ensures converged spectra up to 5 eV, and a numerical uncertainty of less than 10~meV for the binding energy of the lowest-energy excitons. Atomic structures and isosurfaces are visualized using the software package VESTA~\cite{momm-izum11jacr}.

 \subsection*{EELS measurements}

Plan-view transmission-electron microscopy (TEM) samples oriented along the [100] lattice direction of the BaSnO$_{3}$ single crystal are prepared by classical tripod polishing, {\it i.e.}, the same technique and growth conditions as reported in Ref.~\cite{galazka+16jpcm}. Additional smoothing of the surface is obtained by chemical-mechanical polishing (CMP) using the SiO$_2$ suspension. Argon ion-milling is done at liquid nitrogen temperature using a Precision Ion Polishing System (PIPS) at beam energies of 3.5-0.2 keV.

EELS experiments in the scanning transmission electron microscope (STEM-EELS) are carried out using a Nion HERMES microscope. This instrument is equipped with an aberration corrector, a cold-field-emission-gun (FEG), a monochromator at ground potential, and a hybrid-pixel direct-detection camera (Dectris ELA). Such experimental set-up allows one to acquire EELS spectra with energy resolution below 10 meV at 60 kV accelerating voltage. In our case, we use relatively large energy dispersions of 5, 50 and 100 meV/channel, and opt for opening the monochromating slit to allow for more intensity in the beam at the expense of some energy resolution. Using the smallest dispersion and 10 mrad convergence and collection angles, we measure a zero-loss peak with the full width half maximum (ZLP-FWHM) in vacuum of $\sim$40 meV (see Supplementary Figure 6). We monitor the relative alignment of the electron beam and crystalline plane orientation, acquiring spectra using on-axis and off-axis conditions (see Supplementary Figure 4). The on-axis spectra are better suited for the study of the dielectric function using Kramers-Kronig analysis. The off-axis spectra show better statistics and also reproduce well the shape of the signal in the region of the optical gap ($E_{g}^{\mathrm{opt}}$). Finally, only the on-axis spectrum is presented here, but both can be found in the Supplementary Figure 5.

In addition to these measurements, the impact of thickness and surface contamination is studied. On the one hand, we initially find no difference in the apparent optical gap for spectra acquired at thicker or thinner regions. The consistency of these measurements is achieved due to the use of a relatively low beam voltage of 60 kV in our STEM experiments. On the other hand, we find indications that surface contamination, originating from sample preparation and ubiquitous hydrocarbons from the atmosphere, strongly affect our ability to observe a signal from indirect transitions. Thus, the final measurements reported here are performed after the sample is treated using plasma cleaning with Argon. 

The STEM-EELS characterization of BaSnO$_{3}$ is carried out with the objective of measuring low-loss EELS spectra. In this technique, inelastically scattered electrons with energy loss in the range from zero up to a few tens of eV provide an alternative means of studying dielectric properties of thin samples. In contrast to photons, the strongly interacting electrons allow for studying direct and indirect transitions. In particular, the position of the onset of the inelastic signal at a few-eV in a semiconductor material is related to its optical onset. 

 Naked-eye examination of these spectra already hints at the presence of the optical gap at around 3 eV. Indeed, for our experimental set-up, at this energy resolution and owing to the use of a direct-detection camera, background intensity sources are greatly suppressed above 2 eV. We expect the main contributions to this background intensity to follow a power-law model, including the tail of the zero-loss peak and vibrational losses (that are not resolved in this experiment). From a spectrum image acquisition we identify and sum the spectra in a homogeneous area of the specimen. We carefully fit a model to the resulting spectrum that incorporates expressions for the power-law background and band-gap contributions. In our model, signals in the region of an energy gap $E_t$ are represented by the following expression,
\begin{equation}
s(E) =\Theta (E-E_t)\frac{(E-E_t)^r}{1+e^{k(E-E_t)}},
\end{equation}
where $\Theta$ is the Heaviside step function and $r$ is a factor indicating the nature of the transition, either 0.5 or 1.5 for direct or indirect transitions, respectively. In other words, in the former case, $E_t$ is the optical gap. These parameters stem from the expected shape of the EELS signal close to the absorption onset according to the joint density of states (JDOS) and matrix element for single electron transitions between parabolic bands~\cite{Rafferty+98prb}. The sigmoid function fixed by the parameter $k$ is used to represent the decay of the signal far away from the absorption onset where the parabolic approximation does not hold anymore and the JDOS associated to the transition drops. We let the parameter $k$ vary to tune the decay far away from the main absorption onset in order to highlight the weak signal coming from indirect transitions, and fix it to $k$=0 for the direct transitions (no decay). This heuristic is inspired by the behavior of the excitation signal dictated by the Tauc-Lorentz model that is employed in the study of dielectric properties of semiconductor materials~\cite{Eljarrat+19ulm}. Poissonian noise variance of the EELS data is taken into account for the model-based fit~\cite{hyperspy}. This algorithm also calculates the standard deviation for the obtained parameters~\cite{zamani2021}. The presented uncertainties correspond to these values. In some cases, the standard deviation of the model parameters is well below our estimate for the energy resolution. In these cases, we choose to report the ZLP-FWHM value as the measurement uncertainty.

For most materials, the measurement of valence properties such as the band gap using EELS requires considering the impact of various spurious sources, among which relativistic losses are typically most problematic~\cite{Stoeger+08}. BaSnO$_3$ is no exception, and for transmitted electrons at 60 kV accelerating voltage, the Cherenkov threshold $c^2/v^2 \simeq 5 < \mathrm{Re}\varepsilon $ indicates that in this material relativistic losses might contribute to EELS at the vicinity of the band gap and above, {\it i.e.}, $\sim 3-10$ eV. Experimental techniques that aim at reducing or suppressing the impact of bulk relativistic losses (also know as Cherenkov-loss) have to be carefully tailored as they depend on the material in a non-trivial manner. 

Perhaps the most popular method consists of reducing the accelerating voltage aiming to "slow down" the transmitted electrons below the Cherenkov threshold~\cite{kaiser+11, canas+2018}. This threshold is material-specific, and in many cases, it lies well below the voltages that are attainable in most STEM instruments. An alternative method is using off-axis acquisition geometries (sometimes also called dark-field EELS), eluding most of the relativistic losses since these appear at very small scattering angles~\cite{gu+07, korneychuk+18}. These techniques typically require careful experimental set-up, and in most cases convergent-beam electron diffraction (CBED) needs to be considered which is again sample specific.

With the aim to address these issues, we also acquire spectra with larger energy dispersion, up to around 100 eV, including all transitions from the valence bands, some core transitions, and collective excitation of bulk and surface plasmons. In order to study the bulk dielectric properties of BaSnO$_{3}$, we remove plural scattering, using Fourier-log deconvolution and apply relativistic Kramers-Kronig analysis, using an in-house developed algorithm~\cite{Eljarrat+19ulm}. For the latter, we normalize the experimental EELS using a reference refractive-index, $n_{\mathrm{BaSnO}_{3}}$=2.05, achieving convergence after less than 10 iterations. We monitor the convergence by computing the chi-square test between experimental and corrected spectra, with a target value of $X^{2} = 10^{-4}$. This results in an estimate of the complex dielectric function, $\varepsilon(E)$, and of the sample-region thickness of around 120 nm, which is probably underestimated because of beam broadening.

\subsection*{Optical absorption measurements}
Optical transmittance spectra are recorded from 5 K to 300 K using a Lambda 19 (PerkinElmer) spectrophotometer and a liquid helium cryostat (Oxford Instruments). The used sample of 150 $\mu$m thickness is-double side polished. Absorption coefficients are calculated from the transmittance data by approximately taking account of the reflection losses setting the refraction index to 2. More details can also be found in Refs.~\onlinecite{galazka+16jpcm,Aggoune+LIO}.

\section*{Data availability}
Input and output files can be downloaded free of charge from the NOMAD Repository~\cite{drax-sche19jpm} at the following link: \url{https://dx.doi.org/10.17172/NOMAD/2021.05.10-1} (see also Ref.~\onlinecite{note-theory-data}). The EELS and rKKA-EELS data can be found at \url{https://dx.doi.org/10.17172/NOMAD/2021.11.01-1}.

\section*{Code Availability}
The code \exciting~is available at [\url{exciting-code.org}].
\section*{Acknowledgment} 
This work was supported by the project BaStet (Leibniz Senatsausschuss Wettbewerb, No. K74/2017) that is embedded in the framework of GraFOx, a Leibniz ScienceCampus, partially funded by the Leibniz Association. We acknowledge the North-German Supercomputing Alliance (HLRN, project bep00078) for providing HPC resources. A.E., C.K., and C.D. appreciate partial funding by the Deutsche Forschungsgemeinschaft (DFG) - Projektnummer 182087777 - SFB951. W. A. thanks Cecilia Vona for implementing the HSE functional in the \exciting\ code, and Le Fang and Christian Vorwerk for fruitful discussions. A.E. and C.K would like to thank Dr.~Benedikt~Haas for
supporting the electron spectroscopy experiments.

\section*{Additional Information}
Supplementary Information (see below).

\section*{Author contributions}

W.A. performed the \textit{ab initio} calculations and analyzed the data. A.E. and K.I. measured and analyzed the EELS and optical absorption data, respectively; Z.G. grew the single crystals, M.Z. prepared the samples; and all authors discussed the results.
\section*{Competing interests}
The authors declare no competing interests.

\newpage

\renewcommand{\figurename}{}
\renewcommand{\tablename}{}
\setcounter{figure}{0}    
\renewcommand{\thepage}{S\arabic{page}} 
\renewcommand{\thetable}{Supplementary table \arabic{table}}  
\renewcommand{\thefigure}{Supplementary Figure \arabic{figure}} 

\newpage
\onecolumngrid

\newpage
{\centering
{\large 
\textbf{Supporting Information on}\\
\vskip 0.2cm
\textbf{"A consistent picture of excitations in cubic BaSnO$_{3}$ revealed by combining theory and experiment"}}\newline

Wahib Aggoune$^{1,2}$, Alberto Eljarrat$^{1}$, Dmitrii Nabok$^{1,3}$, Klaus Irmscher$^{4}$, Martina Zupancic$^{4}$, Zbigniew Galazka$^{4}$, Martin Albrecht$^{4}$, Christoph Koch$^{1}$, and Claudia Draxl$^{1,2,3}$\\
\textit{$^{1}$Institut f\"{u}r Physik and IRIS Adlershof, Humboldt-Universit\"{a}t zu Berlin, 12489 Berlin, Germany}\\
\textit{$^{2}$Fritz-Haber-Institut der Max-Planck-Gesellschaft, 14195 Berlin, Germany}\\
\textit{$^{2}$European Theoretical Spectroscopy Facility (ETSF)}\\
\textit{$^{4}$Leibniz-Institut f\"{u}r Kristallz\"{u}chtung, 12489 Berlin, Germany}\\

\vskip 0.5cm
\date{today} 
}

 \twocolumngrid
   \normalsize 
 
\section*{Supplementary methods}
\subsection*{Convergence tests}
\begin{figure}
\begin{center}
\includegraphics[width=0.98\columnwidth]{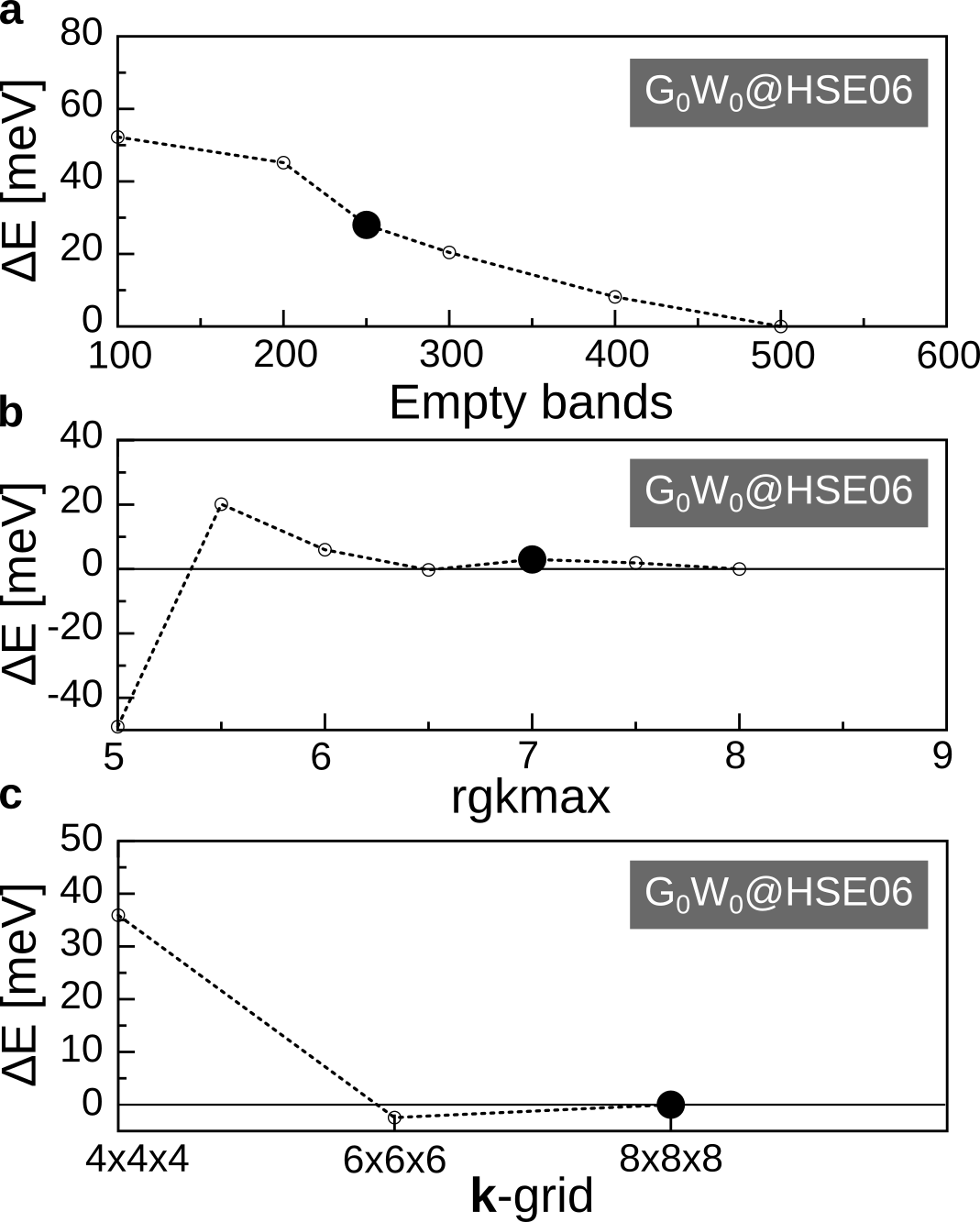}
\caption{\textbf{Convergence tests}. Convergence behavior of the $G_{0}W_{0}$@HSE06 band gap with respect to (a) number of empty bands, (b) cutoff energy (\textit{rgkmax}), and (c) \textbf{k}-grid. The black dots indicates the respective parameter used in the actual calculations shown in the main manuscript.}
\label{fig:S1}
\end{center}
\end{figure}

\begin{figure}
\begin{center}
\includegraphics[width=0.98\columnwidth]{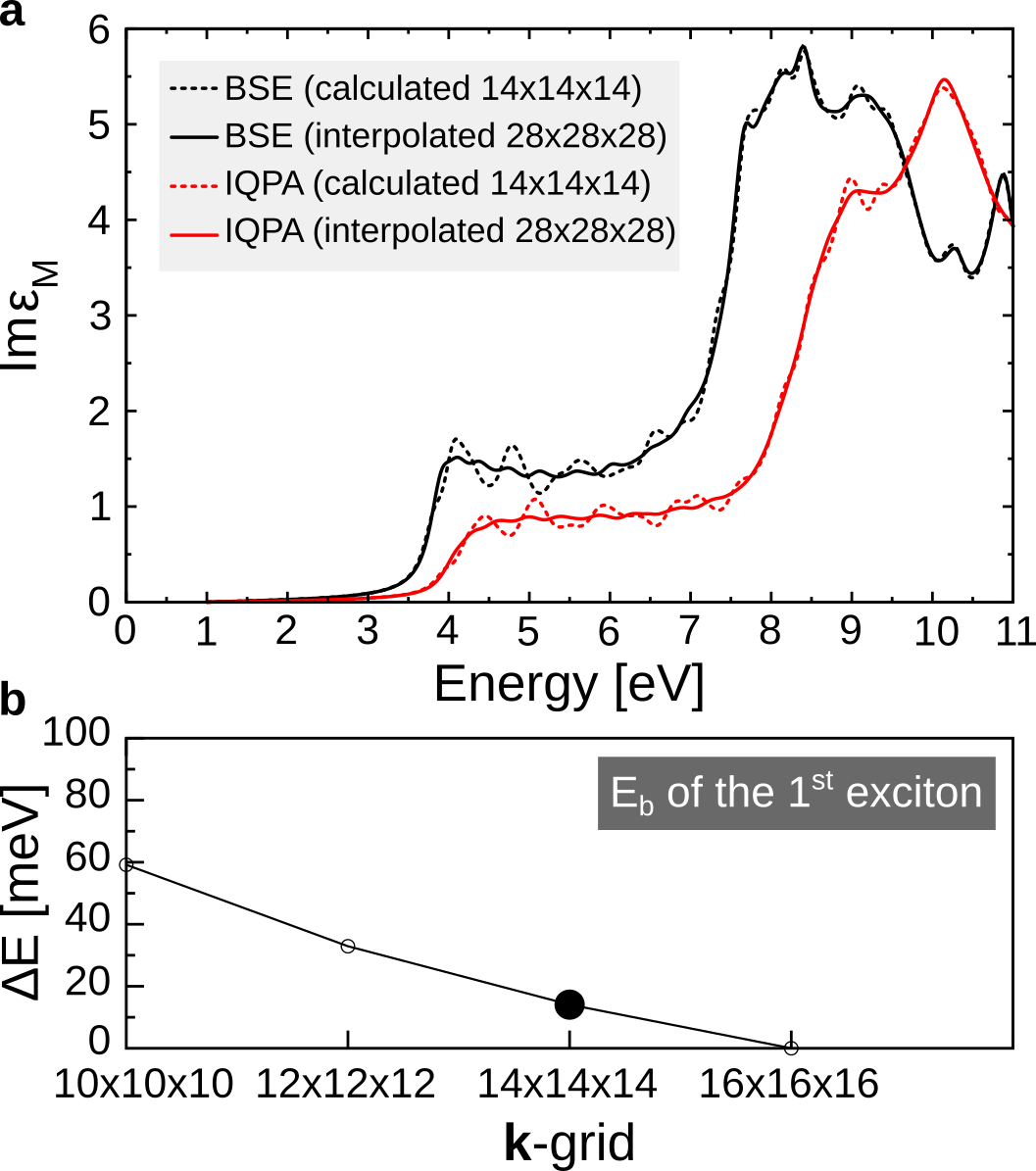}
\caption{\textbf{Convergence behavior of the dielectric function.} (a) Solution of the Bethe-Salpeter equation (BSE, black) for different \textbf{k}-grids. The corresponding spectra obtained within the independent-quasi-particle approximation (IQPA) are shown in red. (b) Convergence of the binding energy of the lowest-energy exciton, obtained by BSE, with respect the \textbf{k}-grid. The black dot indicates the parameter used in the actual calculations.}
\label{fig:S2}
\end{center}
\end{figure}

In~\ref{fig:S1}, we provide convergence tests of the $G_{0}W_{0}$@HSE06 band gap with respect to \textbf{k}-grid, number of empty bands, and basis-set cutoff (in terms of the dimensionless parameter \textit{rgkmax}). We clearly see that a 8$\times$8$\times$8 \textbf{k}-grid and a value of \textit{rgkmax}=7 are good enough to reach a numerical uncertainty of less then 10 meV, while 250 empty bands, as used in our calculations, allows for a precision of about 20 meV. In~\ref{fig:S2}(a), we show the convergence behavior of the absorption spectra. We see that a dense \textbf{k}-grid is needed particularly between 4 and 8 eV. This is mainly due to the fact that this energy range is governed by excitations from the valence band to a small region around the minimum of the highly-dispersive conduction band. For this reason, we adopt a 14$\times$14$\times$14 shifted $\textbf{k}$-point mesh and then apply a double-grid technique~\cite{gille+13prb} to obtain a 28$\times$28$\times$28 mesh. The binding energy of the fist exciton is already converged within 20 meV with a 14$\times$14$\times$14 shifted $\textbf{k}$-grid [panel (b)].

\subsection*{Dielectric function}

\begin{figure}
\begin{center}
\includegraphics[width=0.49\textwidth]{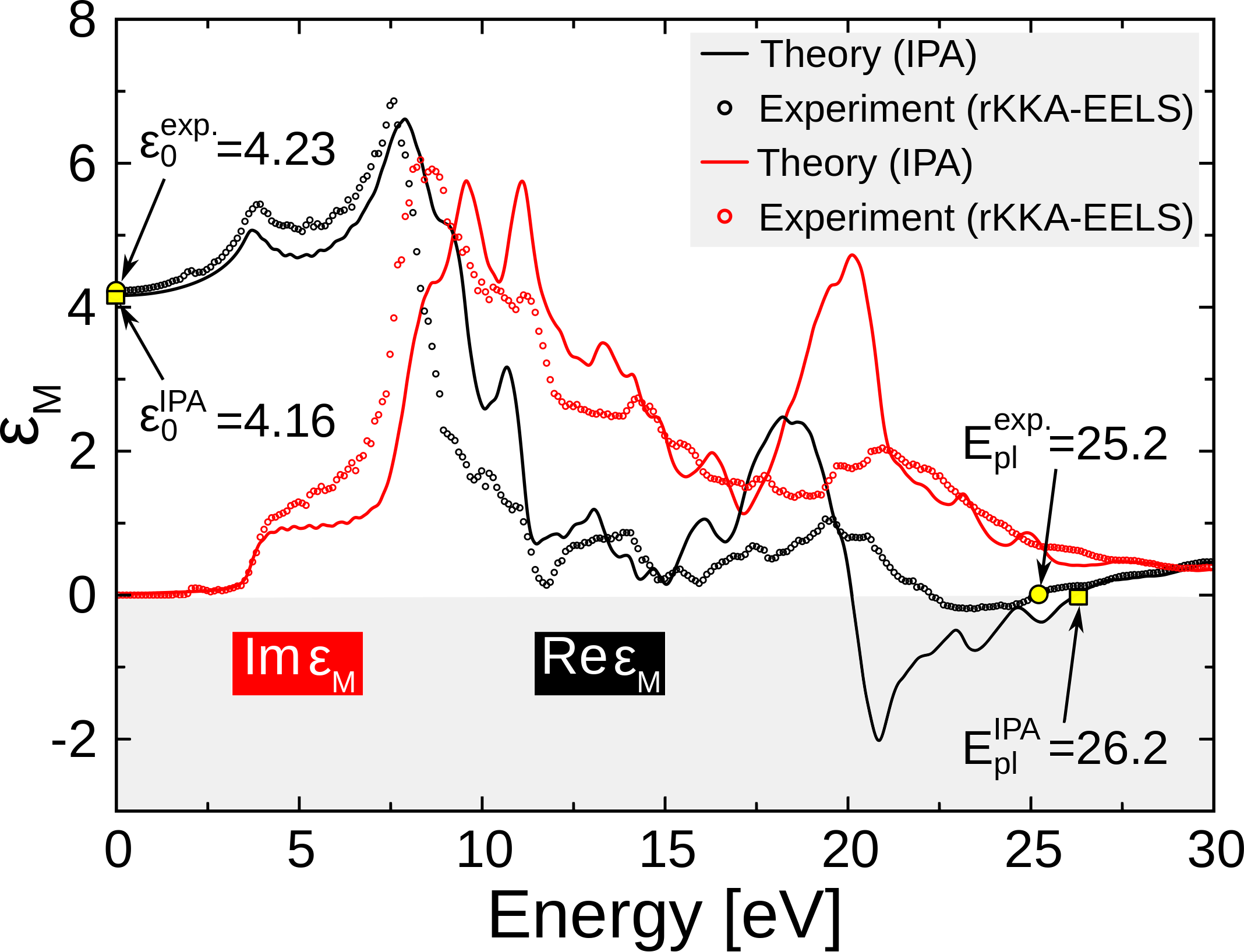}
\caption{\textbf{Dielectric function}. Real (black) and imaginary (red) part of the macroscopic dielectric function of cubic BaSnO$_{3}$, obtained by applying relativistic Kramers-Kronig analysis (rKKA) analysis to the electron energy-loss spectroscopy (EELS) data (circles) and calculated by the independent-quasi-particle approximation (IQPA, solid line). The latter are corrected by -0.367 eV to account for band-gap renormalization due to temperature and zero-point vibrations (ZPV). A Lorentzian broadening of 0.15~eV is applied to mimic excitation lifetimes.}
\label{fig:S3}
\end{center}
\end{figure}

In~\ref{fig:S3}, we depict the real and imaginary part of the microscopic dielectric function obtained by applying relativistic Kramers-Kronig analysis (rKKA) analysis to the electron energy-loss spectroscopy (EELS) data (see Method section in the main text). We clearly see the zero-crossing of the real part from the negative half-plane at about 25.2 eV, which corresponds to the bulk plasmon energy. It is in agreement with an experimental value of 26.1, reported for epitaxially-grown BaSnO$_{3}$ films~\cite{yan+18jvsta}. In the latter, first-principles calculation on the independent-particle level, found another plasmonic peak at 15.2 eV. Such a peak is neither observed in their or our measurements, nor in our calculations.

To this extent, we calculate the dielectric function in the independent particle approximation, applying a scissors shift to the PBEsol band structure to recover the $G_{0}W_{0}$@HSE06 band gap (independent-quasi-particle approximation IQPA), including as well a correction for band-gap renormalization due to temperature and ZPV effects. This calculation is carried out using a shifted 30$\times$30$\times$30 \textbf{k}-grid. A basis-set cutoff G$_{max}$=5 bohr$^{-1}$ and 300 unoccupied states are used to reach a converged spectra, specifically the real part of the dielectric function. The result is shown in~\ref{fig:S3}. Overall, the calculated and measured spectra are in a reasonable agreement, in particular, the values of $\varepsilon_{0}$, obtained as 4.16 eV and 4.23 eV, respectively. The plasmonic peak is found by theory at about 26.2 eV, slightly larger than our measured value of 25.2 eV. For this small discrepancy, both missing excitonic and phonon effects in the former and a simplified model for obtaining the latter can be held responsible.

\begin{figure}
\begin{center}
\includegraphics[width=0.98\columnwidth]{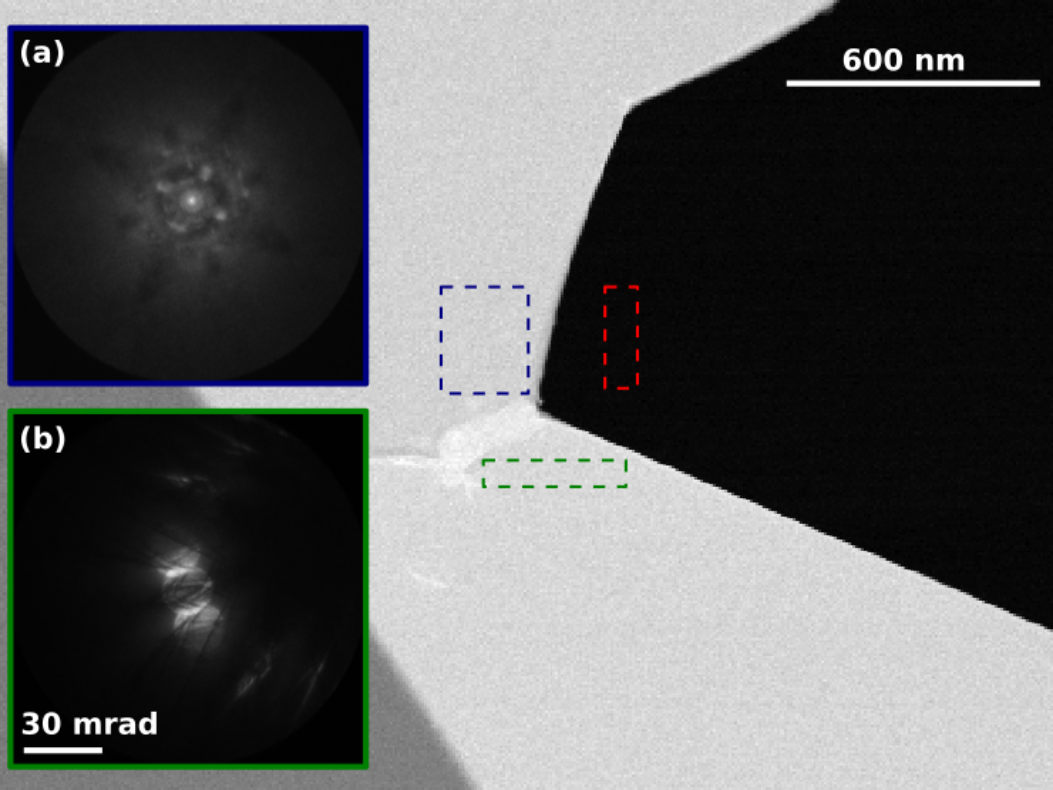}
\caption{\textbf{Analysis of electron-microscopy images}. The main panel shows an high-angle annular dark-field imaging (HAADF) image of the region where on-axis (blue) and off-axis (green) spectrum imaging was performed. Acquired from these regions, convergent-beam electron diffraction (CBED) patterns are depicted in panels (a) and (b), respectively. Exemplary spectra are included in~\ref{fig:S5}. The region where the vacuum reference in~\ref{fig:S6} was acquired is also indicated (red).}
\label{fig:S4}
\end{center}
\end{figure}
\subsection*{Band gap obtained from EELS measurements}
\begin{figure}
\begin{center}
\includegraphics[width=0.48\textwidth]{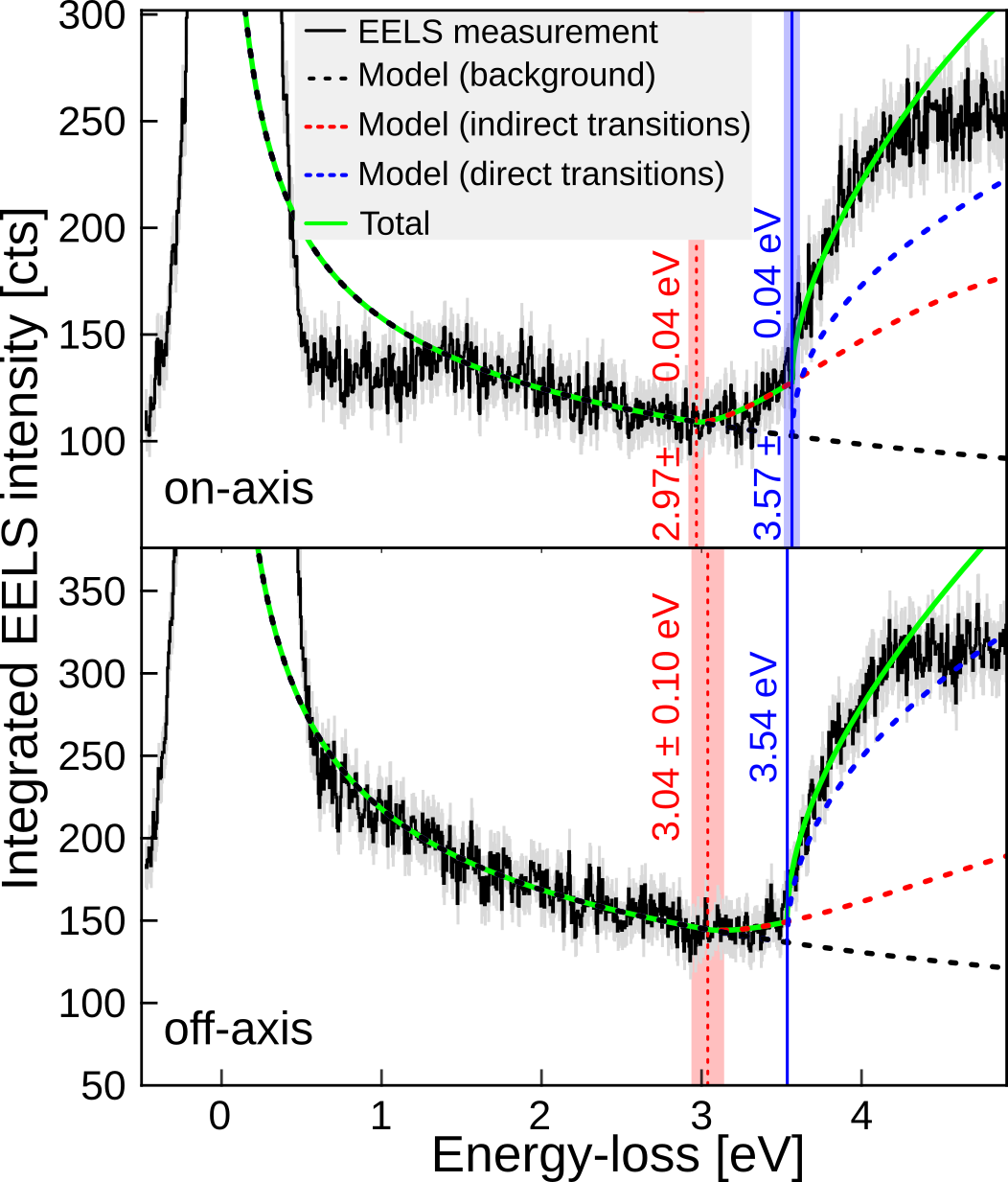}
\caption{\textbf{Band-gap measurements}. Electron energy-loss spectroscopy (EELS) spectra (solid black lines) acquired in on- and off-axis conditions showing the zero-loss-peak (ZLP) tail and band gap signal. The estimated Poisson standard variation is indicated by the gray areas. The fits obtained by a model assuming either indirect or direct transitions (see Method section in the main text) are shown by the red and blue dashed lines, respectively. The dashed black line refers to the background model while the green solid line shows the sum of the three models. The vertical blue line indicates the onset of direct transitions and the dashed red line that of indirect transitions; the shaded areas represents their error bar.}
\label{fig:S5}
\end{center}
\end{figure}

EELS spectra from the regions depicted in \ref{fig:S4} are analyzed to derive the band gap of BaSnO$_3$. Generally speaking, we find that the indirect band gap contributes to the EELS signal rather weakly compared to the direct band gap. This makes it more difficult to distinguish it from the background intensity and the direct-band-gap contributions. We address this issue by averaging over several spectra obtained from a relatively small region in the sample. The resulting averaged spectra (see \ref{fig:S5}) have a higher signal-to-noise ratio (SNR), clearly showing a relatively slow onset at $\sim$3 eV, followed by a more dramatic one at 3.5 eV, corresponding to indirect and direct transitions, respectively.

Poissonian-noise variance of the EELS data is taken into account when performing this averaging and for the model-based fit mentioned in the main text. In this procedure, we employ bounded non-linear least-square fitting using the Levenberg-Marquardt algorithm as implemented in Hyperspy to sequentially fit our model components to the data~\cite{hyperspy}. These components include a power-law background and indirect and direct band gaps. Once this is done, the final parameters are obtained by dual-annealing global optimization also implemented in Hyperspy. This procedure has been used before and is explained more in depth in a previous publication~\cite{zamani2021}.

\begin{figure}[b]
\begin{center}
\includegraphics[width=0.48\textwidth]{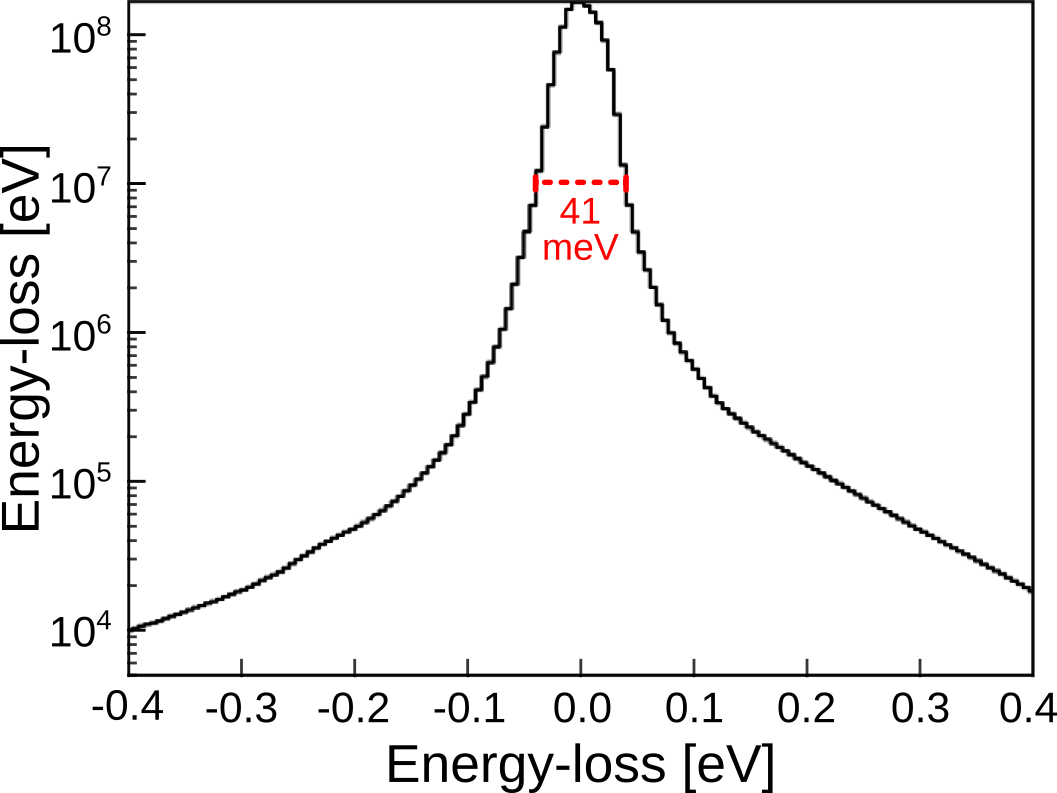}
\caption{\textbf{Zero-loss peak (ZLP)}. Electron energy-loss spectroscopy (EELS) spectrum obtained in vacuum showing the central part of the ZLP. The full width half maximum (FWHM) of this peak, below 50 meV, indicates the energy resolution of our measurement set-up.}
\label{fig:S6}
\end{center}
\end{figure}

This algorithm also calculates the standard deviation for the obtained parameters. The uncertainties presented in the main text and figures such as in \ref{fig:S5} correspond to these values in most cases. In some cases, the standard deviation of the model parameters is well below our estimate for the energy resolution. Here, we choose to report the estimated energy-resolution as the measurement uncertainty.

As depicted in~\ref{fig:S6}, we also measure the zero-loss peak with the full width half maximum (ZLP-FWHM) in vacuum which is around 40 meV for all measurements. This value is taken in the literature as a good indication for the experiment energy-resolution~\cite{egerton+11}.

\end{document}